\documentclass[11pt,a4paper]{scrartcl}

\usepackage{CLICdp}

\usepackage{CLICdp_definitions}

\usepackage{multirow}

\usepackage[normalem]{ulem}

\usepackage{tikz}
\usetikzlibrary{arrows,positioning,shapes}
\usetikzlibrary{decorations.pathmorphing}
\usetikzlibrary{decorations.markings}

\usepackage{cleveref}
\usepackage{tikz}
\usepackage{calrsfs}
\usetikzlibrary{arrows,positioning,shapes}
\usetikzlibrary{decorations.pathmorphing}
\usetikzlibrary{decorations.markings}

\usepackage{cleveref}

\renewcommand{\whizard}{\textsc{Whizard}\xspace}

\DeclareMathAlphabet{\mathcal}{OMS}{cmsy}{m}{n}
\newcommand{\kkmc}{${\cal KK}$\,MC\xspace}

\title{Simulating hard photon production with \whizard}

\nolicence %Uncomment this line to remove the licence comment (not shown for drafts automatically)

\clicdppub{2020}{001}  % journal articles
\date{\today}

\addauthor{J.~Kalinowski}{\institute{1}}
\addauthor{W.~Kotlarski}{\institute{2}}
\addauthor{P.~Sopicki}{\institute{1}}
\addauthor{A.~F.~\.Zarnecki}{\institute{1}}

\addinstitute{1}{Faculty of Physics, University of Warsaw, Poland}
\addinstitute{2}{Institut f\"ur Kern- und Teilchenphysik, TU Dresden}

\abstract{
  One of the important goals of the proposed future $\Pep\Pem$
  collider experiments is the search for dark matter particles
  using different experimental approaches. The most general search
  approach is based on the mono-photon signature, which is expected
  when production of the invisible final state is accompanied by a
  hard photon from initial state radiation. Analysis of the energy
  spectrum and angular distributions of those photons can shed light
  on the nature of dark matter and its interactions. Therefore, it
  is crucial to be able to simulate the signal and background samples
  in a uniform framework, to avoid possible systematic biases. The
  \whizard program is a flexible tool, which is widely used by
  $\Pep\Pem$ collaborations for simulation of many different "new
  physics" scenarios.
  We propose the procedure of merging the matrix element calculations
  with the lepton ISR structure function implemented in \whizard.
  It allows us to reliably simulate the mono-photon events, including
  the two main Standard Model background processes: radiative neutrino pair
  production and radiative Bhabha scattering.
  We demonstrate that cross sections and kinematic distributions of
  mono-photon in neutrino pair-production events agree with
  corresponding predictions of the \kkmc, a Monte Carlo generator
  providing perturbative predictions for SM and QED processes,
  which has been widely used in the analysis of LEP data.
}

\titlecomment{This work was carried out in the framework of the CLICdp Collaboration}

\lstdefinestyle{mylststyle}{
    numberstyle=\tiny,
    basicstyle=\ttfamily\scriptsize,
    breakatwhitespace=false,         
    breaklines=true,                 
    captionpos=b,                    
    keepspaces=true,                 
    numbers=none,   %left/right/none                    
    numbersep=5pt,                  
    showspaces=false,                
    showstringspaces=false,
    showtabs=false,                  
    tabsize=2,
    prebreak=\raisebox{0ex}[0ex][0ex]{\ensuremath{\,\,\hookleftarrow}}, 
    frame=none % none/leftline/topline/bottomline/lines/single/shadowbox
}
 
\graphicspath{{./plots/}}

\begin{document}

\titlepage

% -*- mode: LaTeX; mode: auto-fill; mode: flyspell-prog; -*-

\section{Introduction}

The search for dark matter (DM) particles is one of the main research
goals of many running and planned experiments.
The evidence for the existence of DM has so far been based
only on observations of its gravitational interactions.
Still, it is expected that it can also be produced at colliders, interact
with ordinary matter in direct search detectors or produce signal
detectable in indirect search experiments when annihilating in the
dense regions of the universe.
Many different theoretical models have been proposed to describe the
nature of DM and they result also in many different discovery
scenarios.
Collider experiments, assuming DM particles can be produced at high
energy collisions, have to rely on indirect signatures, as direct observation
of DM particle in the detector is not possible.
Experimental searches are based on the  processes in which the DM
particle production is associated (due to production mechanism or in
the decay chain of more massive objects) with the production of
particular final state objects like jets, massive gauge bosons or
high energy photons.

\begin{figure}[hbp]
  \centerline{\includegraphics[width=0.35\textwidth]{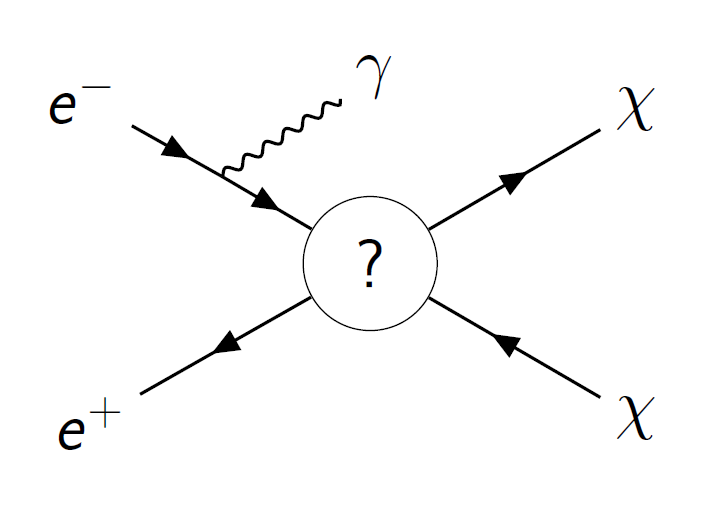}}
  \caption{Diagram describing DM particle pair production process
    with additional ISR photon radiation.}
  \label{fig:sigdiag}
\end{figure}

The mono-photon signature is one of the considered scenarios to look for DM
particle production in future $\Pep\Pem$ colliders.
We usually assume that DM particles, denoted $\PGc$ in the following,
can be pair produced in the $\Pep\Pem$ collisions via exchange of a new
mediator particle, which couples to both Standard Model (SM) particles
and DM states:

\[ \Pep \Pem \to \PGc \; \PGc \; . \]
However, if this is the case, the produced final state is invisible in
the detector.
The simplest way to detect this process is via the observation of
additional hard photon radiation from initial state leptons, as shown
in Fig.~\ref{fig:sigdiag}.
Production of the initial state radiation (ISR) photon should be
independent on the details of the DM production model.
By studying the distribution of photons emitted in the process

\[ \Pep \Pem \to \PGc \; \PGc \; \PGg \]
we should be able to constrain the DM particle production cross
section.
Bounds on DM production processes with the mono-photon signature have been
derived from LEP results
\cite{Achard:2003tx,Abdallah:2003np,Abdallah:2008aa,Fox:2011fx}.
Production of DM particles in this channel has also been
recently considered for ILC running at 500\,GeV \cite{Habermehl:2020njb}.
For proper estimate of the experimental sensitivity,
precise modelling of all background processes is required.

The \whizard program \cite{Kilian:2007gr,Moretti:2001zz} is a flexible
tool, which is widely used for numerical cross section calculations
and generation of Monte Carlo event samples in
collider studies for Standard Model processes and
many different Beyond Standard Model (BSM) scenarios.
The main goal of the presented study is to develop the proper
framework for simulation of SM background processes to mono-photon
study in \whizard.
Main SM background contributions are expected to come from the
radiative neutrino pair production process

\[ \Pep \Pem \to \PGn \; \PGn \; \PGg \]
shown in Fig.~\ref{fig:vvdiag},
which can not be distinguished from the signal process on the detector
level, and the radiative Bhabha scattering

\[ \Pep \Pem \to \Pep \Pem \; \PGg \]
which contribute to the mono-photon background when both electrons
escape undetected along the beam pipe \cite{Habermehl:2020njb}.

The standard procedure to take ISR effects into account when
generating events with \whizard is to use the built-in lepton ISR
structure function which includes all orders of soft and soft-collinear
photons as well as up to the third order in high-energy collinear photons.
However, this approach allows only for a proper modelling of the
kinematics of the hard scattering, but is not suitable when we expect
photons to be detected in the experiment.
The ISR photons generated by \whizard should not be considered as
ordinary final state particles.
Their energy and transverse momenta correspond to the sum over all
photons radiated in the event from a given lepton line.
For proper description of the photon measurement, the hard non-collinear photon
emission should be included in the generation of the considered
background process on the matrix element level.
Generator-level cuts can be applied, corresponding to the detector
acceptance, on the final state photon(s), which should also allow to
remove divergences in the cross section calculations.
%
% \comment{Does it imply we use acceptance cuts on all ME photons?}
%
To avoid double-counting,
a dedicated merging procedure is then used to remove events with
photons from ISR structure function emitted in the same kinematic
region.

We present results obtained for the (radiative) neutrino pair production
and for the (radiative) Bhabha scattering, for the energy range from
240\,GeV to 3\,TeV,
corresponding to the energy reach of the considered future $\Pep\Pem$ collider
projects: CEPC~\cite{CEPCStudyGroup:2018ghi},
FCCee~\cite{Abada:2019lih}, ILC~\cite{Bambade:2019fyw} and
CLIC~\cite{Charles:2018vfv}.
%
%corresponding to the considered future $\Pep\Pem$ collider projects.
%
The proposed approach is more general and can be applied to any
process with photon radiation, which can be simulated in \whizard.
Conceptually the procedure is reminiscent of the MLM matching in
QCD~\cite{Mangano:2006rw}, with simplifications of not including a QED
shower, an $\alpha_{em}$ running, $\gamma^* \to f \bar f$ splittings
in extra emissions or photons in the initial state, which are
justified by the smallness of the coupling, $\alpha_{em}\ll 1$.

% -*- mode: LaTeX; mode: auto-fill; mode: flyspell-prog; -*-

\xspace % needed for using particle names in section title (?!)

\section{Hard photons from $\PGn\PAGn$ events}

\subsection{$\PGn\PAGn$ production}

\begin{figure}[tbp]
  \hspace{0.02\textwidth}
  \includegraphics[width=0.3\textwidth]{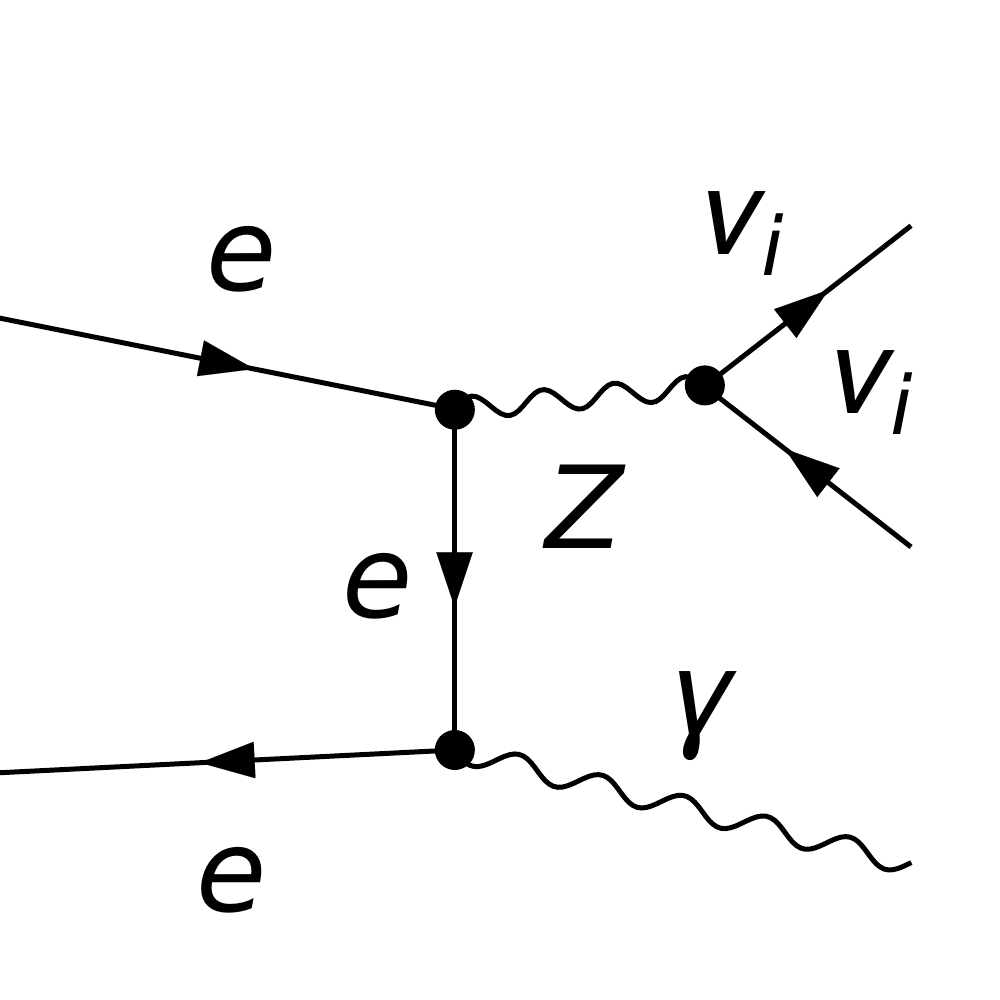}
  \hspace{0.02\textwidth}
  \includegraphics[width=0.3\textwidth]{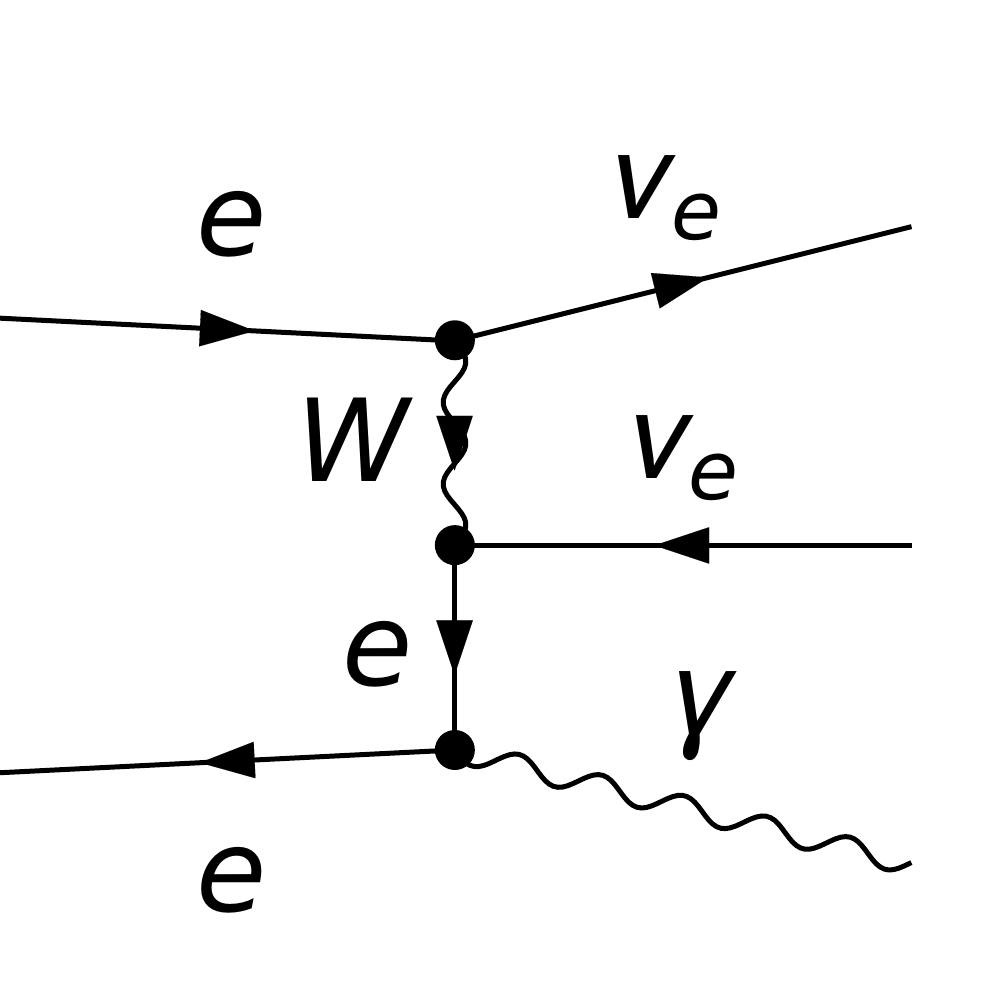}
  \hspace{0.02\textwidth}
  \includegraphics[width=0.3\textwidth]{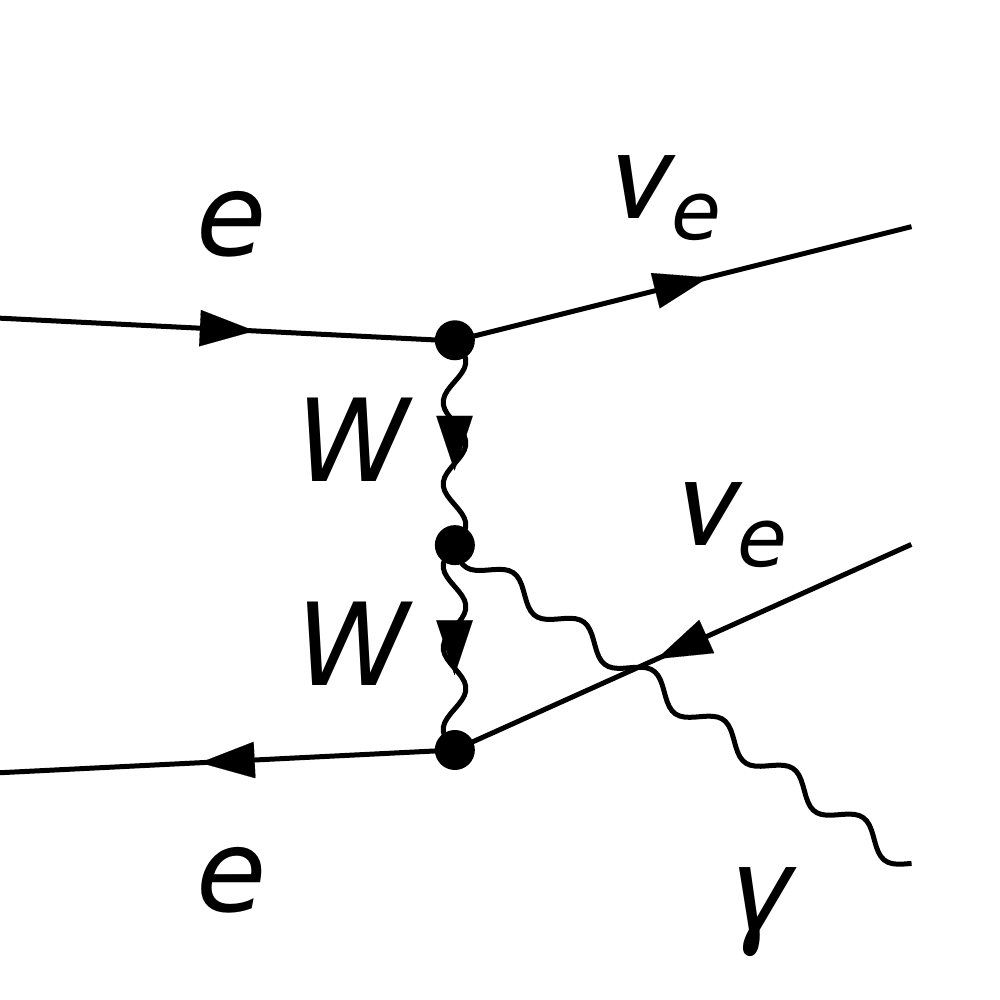}
  \caption{Diagrams describing the neutrino pair production process
    with additional photon radiation.}
  \label{fig:vvdiag}
\end{figure}

Two SM diagrams contribute to the neutrino pair production in $\Pep\Pem$
collisions: the one with s-channel $\PZ$ boson production and
t-channel exchange of the $\PWpm$ boson.
For both processes, the ISR photon can be emitted by the incoming
electron or positron, but for the process with the $\PWpm$ boson
exchange, radiation is also possible from the $\PWpm$ line, see
Fig.~\ref{fig:vvdiag}.
This is an additional argument to conclude that the standard approach
to ISR photon generation, as implemented in \whizard, should not be used
for detailed simulation of radiative neutrino pair production events.
Contribution of the diagram with photon radiation from the $\PWpm$
line significantly reduces the cross sections for radiative neutrino
pair production\footnote{This was verified by switching
  off the $\PWp\PWm\PGg$ vertex in \whizard process definition. Note
  that the results obtained without this vertex are gauge dependent.}
with the largest impact observed for multi-photon
events and events with high photon transverse momenta.
Therefore, for precise kinematic description of photons entering the
detector, we need to include hard photon emission directly in the
process matrix element (ME) calculation.
On the other hand, very soft and collinear photons should still be
simulated with the parametric approach, taking into account proper
summation of higher order corrections.
That is why a dedicated procedure for merging between the two regimes is needed.

%---------------------------------------------------------------------------

\subsection{ME-ISR merging}

\label{sec:detector}

In the presented study we use the following variables, calculated
separately for each emitted photon,  to describe
kinematics of the emission:
\begin{eqnarray*}
  q_{-} & = & \sqrt{4 E_{_0} E_\gamma} \cdot
  \sin{\frac{\theta_\gamma}{2}} \; , \\
  q_{+} & = & \sqrt{4 E_{_0} E_\gamma} \cdot
  \cos{\frac{\theta_\gamma}{2}} \; ,
  \end{eqnarray*}
where $E_{_0}$ is the nominal $\Pep$ and $\Pem$ beam energy, while $E_\gamma$
and $\theta_\gamma$ are the energy and scattering angle of the emitted
photon in question.
For the single photon emission they would correspond to the virtuality
of the electron or positron after (real) photon emission.

\begin{figure}[tbp]
  \includegraphics[width=0.49\textwidth]{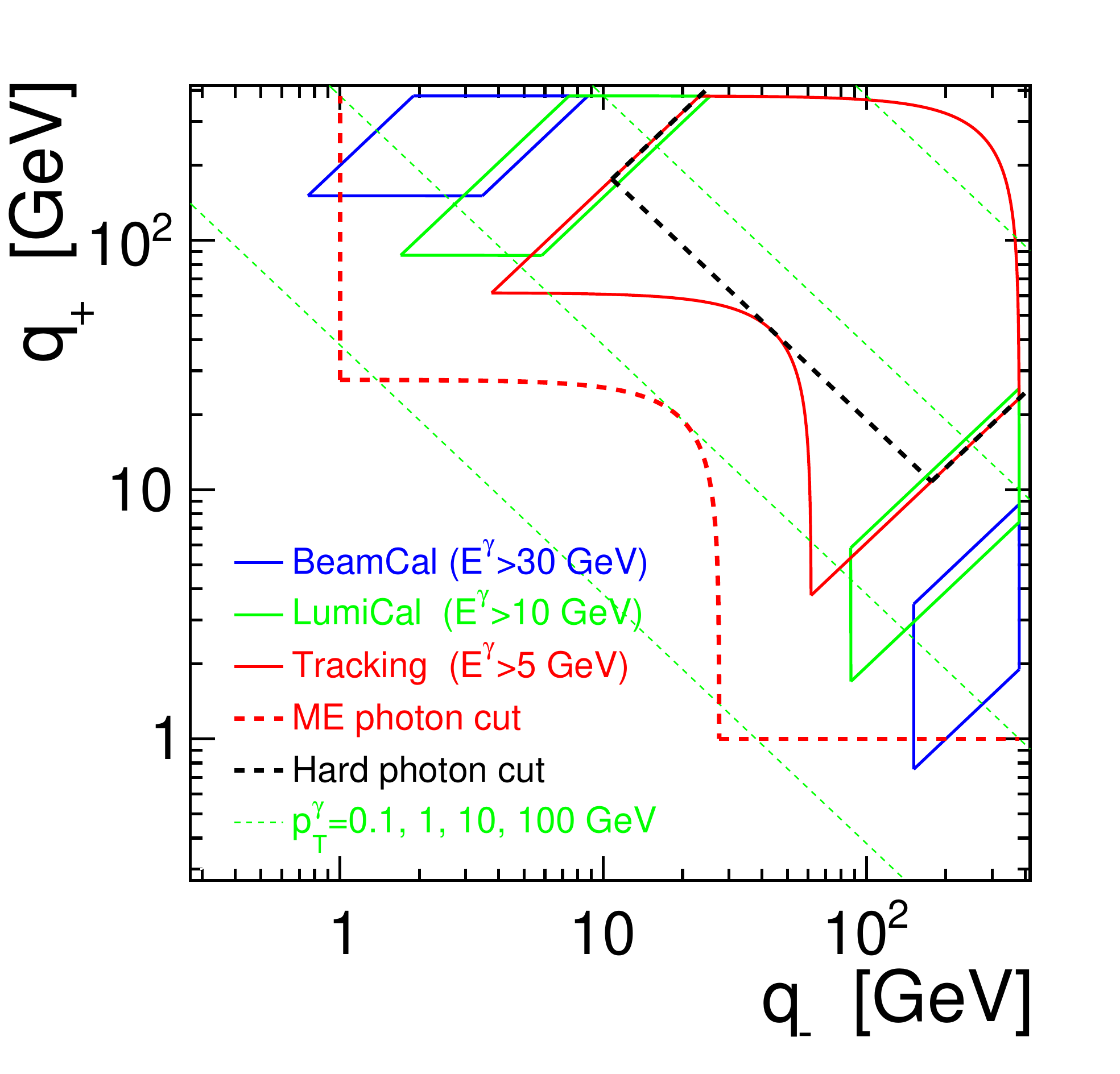}
  \includegraphics[width=0.49\textwidth]{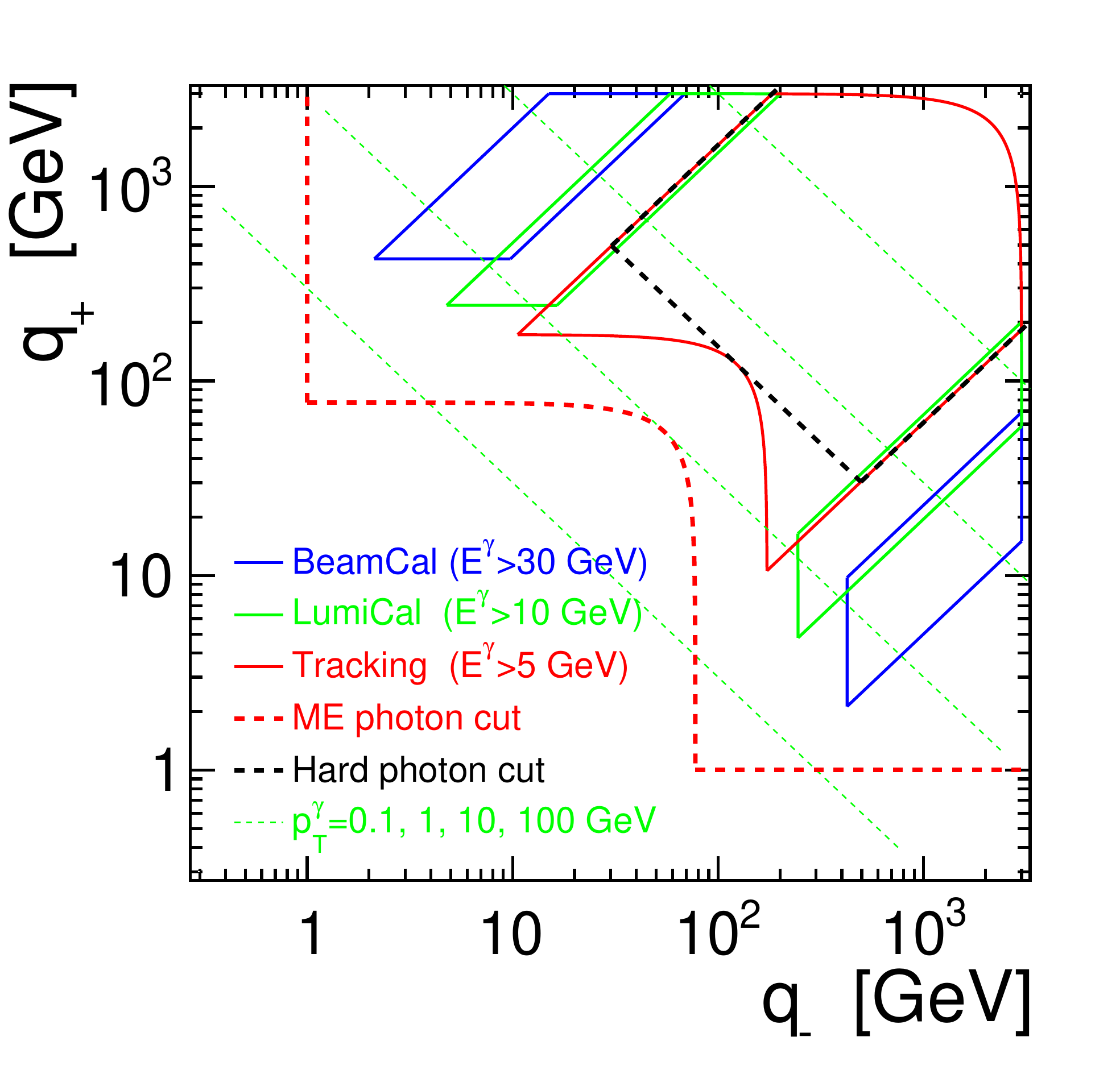}
  \caption{Detector acceptance expected for the future experiments
    at $\Pep\Pem$ colliders in the $(q_+,q_-)$ plane, for collision energies
    of 380\,GeV (left) and 3\,TeV (right). Dashed lines indicate a default cut
    used to restrict the phase space for ME photon generation (red)
    and the cut used for hard photon selection (black), see text for details.}
  \label{fig:q_plot}
\end{figure}

Variables $q_{-}$ and $q_{+}$ are independent and the pair of values
($q_-$,$q_+$) gives the information on both the energy and scattering
angle of a given photon.
Shown in Fig.~\ref{fig:q_plot} is the expected CLIC detector coverage
in the ($q_-$,$q_+$) space, for 380\,GeV and 3\,TeV running.
We assume that photons will be well reconstructed in the detector
for scattering angles between 7$^\circ$ and 173$^\circ$ (corresponding
to the full efficiency of the central tracking detectors, required to
reject electron background) and above the energy threshold of 5\,GeV.
Also indicated are the expected acceptances of the beam calorimeter
(BeamCal: angular coverage from 10 to 46 mrad, energy threshold of
30\,GeV) and luminosity calorimeter (LumiCal: angular coverage 39 to
134 mrad, energy threshold of 10 GeV) \cite{clicdet}.
Only photons with large values of virtualities $q_-$ and $q_+$ can be
measured in the detector.
We therefore require photons generated at the ME level to have energy
above $E_{min}=1$\,GeV and $q_\pm$ virtualities above the merging scale
% $q_{min}=1$\,GeV, as indicated with red dashed line in  Fig.~\ref{fig:q_plot}.
$q_{min}$.
ISR is also taken into account in the cross section integration and
generating events, always resulting in two additional photons in the
event.\footnote{\whizard always puts the two ISR photons first in the
  event record. This allows for easy separation of ME and ISR photons
  in \whizard cuts and event selection, see Appendix~\ref{sec:a1} for
  an example of Sindarin scripts.}
The transverse momenta of ISR photons are taken into account on the
event simulation level.\footnote{Implemented ISR handler is used with
  parameter \texttt{\$isr\_handler\_mode = ``recoil''}.}
At the same time, to avoid double counting, we reject the events with
any of the ISR photons passing the ME photon selection
cuts.\footnote{The \whizard parameter \texttt{isr\_q\_max} should in
  principle allow to limit the kinematic range of the ISR emission.
  However, we noticed that it does not affect the cross section
  integration so the calculated cross section values can not be used
  for normalisation of the generated event samples. That is why we
  have to use the ISR rejection procedure in \whizard instead.}
This procedure will be referred to as 'ISR rejection' in the following.
After ISR rejection, the phase space for photon radiation is
unambiguously divided into ME emission and ISR regions, as
indicated with red dashed line in Fig.~\ref{fig:q_plot} for merging
scale $q_{min}=1$\,GeV.
As mentioned above, photons generated by \whizard from the 
ISR structure function should not be considered as single physical
particles, but correspond to the sum over all  photons radiated in the
event from a given lepton line.
Therefore, the proposed merging procedure is only approximate.
 
For the results presented in this paper, version 2.8.2 of
\whizard \cite{Kilian:2007gr,Moretti:2001zz} was used, compiled with
extended precision option.\footnote{Extended precision was crucial for
 the convergence of the Bhabha cross section integration at
 highest collision energies.} 
Shown in Tab.~\ref{tab:vv_tab_Escan} are the cross sections for
neutrino pair production, for different multiplicities of ME photons
(up to three) and different collision energies.
The statistical uncertainties resulting from the \whizard integration
are below or of the order of one permille for processes without or with
one ME photon and increase up to 5\% for processes with three photons.
The second column gives the cross section for neutrino pair production
with ISR photon radiation (before ISR rejection).
The effect of the ISR is largest at low collision energies, where it
increases the cross section by up to 25\% for 240\,GeV.
The contribution from the 'radiative return' results in the increase
of the cross section for the lowest energies.
The ISR contribution becomes negligible at the highest collision
energies, where the neutrino pair production process is dominated by
the $\PWpm$ exchange diagram.

The last column in Tab.~\ref{tab:vv_tab_Escan} gives the total cross section (sum over photon
multiplicities) expected after ISR-ME merging, i.e. after removing
events with ISR photons passing the ME photon selection (ISR rejection).
The total cross section after ISR-ME merging is very close to the
cross section for the $\Pep\Pem\rightarrow\PGn\PAGn$ process with ISR generation
only (without ISR rejection) showing that the merging procedure does
not affect the normalisation on the sample.
\begin{table}[tbp]
\centering
{\scriptsize
  \begin{tabular}{||c|c|c|c|c|c||}
\hline
\multirow{2}{4em}{$\sqrt{s}[GeV]$} & 
	\multicolumn{4}{c|}{ Whizard-2.8  $\sigma(\Pep\Pem\rightarrow\PGn\PAGn)$ [fb] } &
	\multirow{2}{8em}{$\sigma(\Pep\Pem\rightarrow\PGn\PAGn)$ [fb] after ISR rejection} \\ 
 & $\PGn\PAGn$ &  $\PGn\PAGn+\gamma_{ME}$ &  $\PGn\PAGn+2\gamma_{ME}$ &  $\PGn\PAGn+3\gamma_{ME}$ & \\ 
\hline
240& 53900  \iffalse 53940.19 +- 42.1 \fi & 12600  \iffalse 12591.954 +- 15.1 \fi & 1300  \iffalse 1336.6096 +- 3.02 \fi & 64  \iffalse 63.926145 +- 3.64 \fi & 53800  \iffalse 67932.68*0.79255 +- 45.0 \fi \\ 
250& 53300  \iffalse 53298.577 +- 40.5 \fi & 12200  \iffalse 12233.589 +- 14.9 \fi & 1300  \iffalse 1297.3419 +- 2.82 \fi & 60  \iffalse 60.074963 +- 2.96 \fi & 53000  \iffalse 66889.583*0.79255 +- 43.3 \fi \\ 
380& 50900  \iffalse 50887.804 +- 40.9 \fi & 10900  \iffalse 10928.712 +- 13.3 \fi & 1200  \iffalse 1159.1073 +- 2.76 \fi & 55  \iffalse 54.900378 +- 4.42 \fi & 50600  \iffalse 63030.524*0.8023 +- 43.4 \fi \\ 
500& 51200  \iffalse 51154.507 +- 31.7 \fi & 11300  \iffalse 11270.64 +- 14.4 \fi & 1200  \iffalse 1232.556 +- 3.54 \fi & 75  \iffalse 74.725207 +- 3.38 \fi & 51200  \iffalse 63732.428*0.80395 +- 35.2 \fi \\ 
1000& 52800  \iffalse 52784.907 +- 43.7 \fi & 13600  \iffalse 13615.835 +- 18.3 \fi & 1700  \iffalse 1749.3116 +- 6.33 \fi & 120  \iffalse 119.46851 +- 4.98 \fi & 52200  \iffalse 68269.522*0.7642 +- 48.1 \fi \\ 
1500& 53300  \iffalse 53339.396 +- 40.6 \fi & 15400  \iffalse 15357.677 +- 22.9 \fi & 2200  \iffalse 2181.8137 +- 7.51 \fi & 170  \iffalse 172.67074 +- 5.82 \fi & 52100  \iffalse 71051.557*0.7328 +- 47.6 \fi \\ 
3000& 53900  \iffalse 53936.615 +- 41.6 \fi & 18500  \iffalse 18450.962 +- 27.7 \fi & 3100  \iffalse 3136.8783 +- 8.93 \fi & 180  \iffalse 176.22857 +- 12.9 \fi & 52400  \iffalse 75700.684*0.6918 +- 52.4 \fi \\ 
\hline 
\end{tabular} 

}
\caption{Cross section values for (radiative) $\PGn\PAGn$ events with
  different collision energies and different multiplicities of photons
  included in the matrix element calculations.
  Number of significant digits indicates the statistical
    precision resulting from \whizard integration.
}
\label{tab:vv_tab_Escan}
\end{table}
Shown in Tab.~\ref{tab:vv_tab_qscan} are the cross section values for
two selected collision energies, 380\,GeV and 3\,TeV, but for
different merging scales $q_{min}$.
\begin{table}[tbp]
\centering
{\scriptsize
    \begin{tabular}{||c|c|c|c|c|c|c||}
\hline
\multirow{2}{4em}{$\sqrt{s} [GeV]$} & \multirow{2}{4em}{$q_{min}[GeV]$} &
\multicolumn{4}{c|}{ Whizard-2.8 $\sigma(\Pep\Pem\rightarrow\PGn\PAGn)$ [fb]} &
\multirow{2}{8em}{$\sigma(\Pep\Pem\rightarrow\PGn\PAGn)$ [fb] after ISR rejection} \\ 
& & $\PGn\PAGn$ &  $\PGn\PAGn+\gamma_{ME}$ &  $\PGn\PAGn+2\gamma_{ME}$ & $\PGn\PAGn+3\gamma_{ME}$ &\\ 
\hline
\multirow{3}{4em}{380} 
& $q_{min}$=0.1 &   \iffalse 50834.491 +- 33.5 \fi & 16500  \iffalse 16528.605 +- 19.1 \fi & 2600  \iffalse 2647.6678 +- 5.23 \fi & 220  \iffalse 221.33099 +- 10.2 \fi & 50000  \iffalse 70232.094*0.7116 +- 40.2 \fi \\ 
& $q_{min}$=0.5 &   \iffalse 50842.873 +- 41.1 \fi & 12600  \iffalse 12609.864 +- 15.3 \fi & 1600  \iffalse 1551.9791 +- 3.69 \fi & 110  \iffalse 107.85391 +- 3.78 \fi & 50400  \iffalse 65112.569*0.7748 +- 44.2 \fi \\ 
& $q_{min}$=1 & 50900  \iffalse 50887.804 +- 40.9 \fi & 10900  \iffalse 10928.712 +- 13.3 \fi & 1200  \iffalse 1159.1073 +- 2.76 \fi & 55  \iffalse 54.900378 +- 4.42 \fi & 50600  \iffalse 63030.524*0.8023 +- 43.4 \fi \\ 
& $q_{min}$=5 &   \iffalse 50841.435 +- 36.3 \fi & 7000  \iffalse 7006.3434 +- 8.99 \fi & 480  \iffalse 479.76063 +- 1.3 \fi & 15  \iffalse 14.862794 +- 0.8 \fi & 50700  \iffalse 58342.402*0.869 +- 37.4 \fi \\ 
& $q_{min}$=10 &   \iffalse 50902.94 +- 38.0 \fi & 5300  \iffalse 5283.046 +- 7.0 \fi & 270  \iffalse 273.88798 +- 1.02 \fi & 7  \iffalse 6.5307226 +- 0.341 \fi & 50800  \iffalse 56466.405*0.9004 +- 38.7 \fi \\ 
& $q_{min}$=50 &   \iffalse 50785.447 +- 30.7 \fi & 1400  \iffalse 1426.5323 +- 1.85 \fi & 21  \iffalse 21.061471 +- 0.319 \fi & 0  \iffalse 0.1705176 +- 0.00899 \fi & 50500  \iffalse 52233.211*0.9676 +- 30.7 \fi \\ 
\hline 
\multirow{3}{4em}{3000} 
& $q_{min}$=0.1 &   \iffalse 53844.509 +- 29.9 \fi & 26200  \iffalse 26174.766 +- 37.2 \fi & 6300  \iffalse 6311.2383 +- 16.0 \fi & 970  \iffalse 965.54418 +- 8.34 \fi & 52200  \iffalse 87296.058*0.5978 +- 51.0 \fi \\ 
& $q_{min}$=0.5 &   \iffalse 53931.639 +- 30.8 \fi & 20700  \iffalse 20731.115 +- 31.4 \fi & 4000  \iffalse 3954.4729 +- 10.9 \fi & 410  \iffalse 411.21673 +- 18.5 \fi & 52300  \iffalse 79028.444*0.66195 +- 49.0 \fi \\ 
& $q_{min}$=1 & 53900  \iffalse 53936.615 +- 41.6 \fi & 18500  \iffalse 18450.962 +- 27.7 \fi & 3100  \iffalse 3136.8783 +- 8.93 \fi & 180  \iffalse 176.22857 +- 12.9 \fi & 52400  \iffalse 75700.684*0.6918 +- 52.4 \fi \\ 
& $q_{min}$=5 &   \iffalse 53955.425 +- 41.0 \fi & 13000  \iffalse 13018.055 +- 20.5 \fi & 1600  \iffalse 1571.1887 +- 8.16 \fi & 74  \iffalse 74.170753 +- 5.98 \fi & 52700  \iffalse 68618.84*0.76765 +- 47.0 \fi \\ 
& $q_{min}$=10 &   \iffalse 53883.099 +- 33.6 \fi & 10700  \iffalse 10679.637 +- 17.8 \fi & 1100  \iffalse 1067.7361 +- 5.87 \fi & 57  \iffalse 57.45971 +- 4.89 \fi & 52600  \iffalse 65687.931*0.80045 +- 38.8 \fi \\ 
& $q_{min}$=50 &  \iffalse 53977.526 +- 38.6 \fi & 5200  \iffalse 5234.1558 +- 9.57 \fi & 260  \iffalse 259.53925 +- 1.59 \fi & 2  \iffalse 2.327644 +- 0.181 \fi & 52800  \iffalse 59473.549*0.8872 +- 39.8 \fi \\ 
\hline 
\end{tabular} 

}

\caption{Cross section values for different merging parameter
  $q_{min}$  in (radiative) $\PGn\PAGn$ events for different multiplicities of photons
  included in the matrix element calculations.
  Number of significant digits indicates the statistical
    precision resulting from \whizard integration.
}
\label{tab:vv_tab_qscan}
\end{table}
Cross sections for different photon multiplicities strongly depend on
the  $q_{min}$ parameter.
However, the total cross section after ISR rejection is basically
independent of this parameter.
We take $q_{min} = 1$\,GeV as the default value for the merging
parameter in the following.

%---------------------------------------------------------------------------

\subsection{Hard photon selection}

Only for a small fraction of neutrino pair production events the
radiated photon can be measured in the detector.
Therefore, additional photon selection on the generator level is required
for efficient simulation of mono-photon events.
We assume the final signal selection will require photon to be reconstructed
in the angular range  $7^\circ < \theta^\gamma < 173^\circ$ and with the
transverse momentum $p_{T}^\gamma > p_{T}^{min} = 5$\,GeV.
These requirements will be referred to as 'hard photon selection' in
the following and ME photons passing this selection will be described
as 'hard'.
For multi-photon events, at least one ME photon needs to pass the hard
photon selection.
An example of the corresponding \whizard steering file is given in
Appendix~\ref{sec:a1}.

Shown in Tab.~\ref{tab:vv_tab_Escan_hard} are the cross sections for
neutrino pair production with hard photon emission, for different
multiplicities of ME photons and different collision energies.
The cross sections listed for different photon multiplicities do not
include ISR photon rejection requirement. The total cross section
expected after the ISR-ME merging procedure is listed in the last
column.

\begin{table}[tbp]
\centering
{\scriptsize
    \begin{tabular}{||c|c|c|c|c||}
\hline
\multirow{2}{4em}{$\sqrt{s}[GeV]$} & 
	\multicolumn{3}{c|}{ Whizard-2.8  $\sigma(\Pep\Pem\rightarrow\PGn\PAGn)$ [fb] } &
	\multirow{2}{8em}{$\sigma(\Pep\Pem\rightarrow\PGn\PAGn)$ [fb] after ISR rejection} \\ 
 &  $\PGn\PAGn+\gamma_{ME}$ &  $\PGn\PAGn+2\gamma_{ME}$ &  $\PGn\PAGn+3\gamma_{ME}$ & \\ 
\hline
240& 4200  \iffalse 4246.8403 +- 5.48 \fi & 730  \iffalse 727.1548 +- 1.63 \fi & 37  \iffalse 37.216 +- 2.39 \fi & 4100  \iffalse 5011.2111*0.82365 +- 6.2 \fi \\ 
250& 4100  \iffalse 4057.2767 +- 5.19 \fi & 700  \iffalse 696.43493 +- 1.56 \fi & 48  \iffalse 48.119938 +- 2.95 \fi & 3900  \iffalse 4801.8316*0.81855 +- 6.17 \fi \\ 
380& 3200  \iffalse 3158.8319 +- 4.24 \fi & 570  \iffalse 572.63636 +- 1.51 \fi & 43  \iffalse 42.536287 +- 1.42 \fi & 3100  \iffalse 3774.0046*0.80975 +- 4.72 \fi \\ 
500& 3100  \iffalse 3113.5474 +- 4.27 \fi & 590  \iffalse 585.4638 +- 1.7 \fi & 48  \iffalse 48.316157 +- 1.4 \fi & 3000  \iffalse 3747.3273*0.801 +- 4.81 \fi \\ 
1000& 3500  \iffalse 3519.6512 +- 5.71 \fi & 780  \iffalse 781.28539 +- 2.98 \fi & 74  \iffalse 73.646737 +- 2.57 \fi & 3300  \iffalse 4374.5833*0.76215 +- 6.93 \fi \\ 
1500& 3800  \iffalse 3815.7606 +- 6.17 \fi & 940  \iffalse 939.29114 +- 4.78 \fi & 98  \iffalse 97.656541 +- 5.52 \fi & 3600  \iffalse 4852.7083*0.7409 +- 9.56 \fi \\ 
3000& 4200  \iffalse 4191.9015 +- 7.48 \fi & 1300  \iffalse 1258.9714 +- 7.08 \fi & 140  \iffalse 142.35464 +- 6.27 \fi & 3900  \iffalse 5593.2276*0.6928 +- 12.1 \fi \\ 
\hline 
\end{tabular} 

}
\caption{Cross section values for (radiative) $\PGn\PAGn$ events, for
  different collision energies and different multiplicities of photons
  included in matrix element calculations with a requirement of at
  least one of the matrix elements photons to be 'hard'.
  Number of significant digits indicates the statistical
    precision resulting from \whizard integration.
}
\label{tab:vv_tab_Escan_hard}
\end{table}

Compared in Tab.~\ref{tab:vv_tab_qscan_hard} are the cross section
values obtained for different merging scales $q_{min}$, for
two selected collision energies, 380\,GeV and 3\,TeV.
Cross section for single hard photon production
($\nu\bar{\nu}+\gamma_{ME}$) is expected to be independent of  $q_{min}$,
as the hard photon selection is more restrictive than the virtuality
cuts applied on ME photons.
Deviation from this expectation is only observed for
$q_{min}=50$\,GeV, when the virtuality cut enters the kinematic region
of the hard photon selection, see Fig.~\ref{fig:q_plot}.

For processes with two or three photons emitted on ME level, the cross
sections  strongly depend on the merging scale, as we require only one
of the photons to pass hard photon selection cuts.
Nevertheless, the total cross section for neutrino pair production
with hard photon radiation, after ISR rejection is applied (last
column in Tab.~\ref{tab:vv_tab_qscan_hard}) is independent of
$q_{min}$.
This confirms that the proposed ISR-ME merging procedure, despite
being only approximate, works very well. 

The influence of the beamstrahlung on the neutrino pair production
turns out to be small.
Taking into account the luminosity spectra expected for CLIC running
at 380\,GeV and 3\,TeV
resulted in 1.5\% and -3.5\% change of the total cross sections
(after ISR rejection), respectively. 
The cross section contribution from the process with four ME photons
in the final state is at per mille level and was neglected in this study.

\begin{table}[tbp]
\centering
{\scriptsize
    \begin{tabular}{||c|c|c|c|c|c||}
\hline
\multirow{2}{4em}{$\sqrt{s} [GeV]$} & \multirow{2}{4em}{$q_{min}[GeV]$} &
	\multicolumn{3}{c|}{ Whizard-2.8 $\sigma(\Pep\Pem\rightarrow\PGn\PAGn)$ [fb] } &
	\multirow{2}{8em}{$\sigma(\Pep\Pem\rightarrow\PGn\PAGn)$ [fb] after ISR rejection} \\ 
& &  $\PGn\PAGn+\gamma_{ME}$ &  $\PGn\PAGn+2\gamma_{ME}$ & $\PGn\PAGn+3\gamma_{ME}$ &\\ 
\hline
\multirow{3}{4em}{380} 
& $q_{min}$=0.1 & 3200  \iffalse 3163.2707 +- 4.33 \fi & 910  \iffalse 910.24515 +- 2.43 \fi & 120  \iffalse 115.58014 +- 4.71 \fi & 3000  \iffalse 4189.0959*0.72535 +- 6.85 \fi \\ 
& $q_{min}$=0.5 & 3200  \iffalse 3160.0425 +- 4.07 \fi & 670  \iffalse 672.65597 +- 1.81 \fi & 55  \iffalse 54.546367 +- 2.92 \fi & 3000  \iffalse 3887.2448*0.78315 +- 5.33 \fi \\ 
& $q_{min}$=1 & 3200  \iffalse 3158.8319 +- 4.24 \fi & 570  \iffalse 572.63636 +- 1.51 \fi & 43  \iffalse 42.536287 +- 1.42 \fi & 3100  \iffalse 3774.0046*0.80975 +- 4.72 \fi \\ 
& $q_{min}$=5 & 3200  \iffalse 3162.1842 +- 4.3 \fi & 340  \iffalse 335.07172 +- 0.874 \fi & 10  \iffalse 9.9661957 +- 0.651 \fi & 3100  \iffalse 3507.2221*0.8723 +- 4.44 \fi \\ 
& $q_{min}$=10 & 3200  \iffalse 3162.9301 +- 4.07 \fi & 230  \iffalse 230.01155 +- 0.762 \fi & 6  \iffalse 6.3513495 +- 0.326 \fi & 3100  \iffalse 3399.293*0.9022 +- 4.16 \fi \\ 
& $q_{min}$=50 & 1400  \iffalse 1427.4315 +- 1.95 \fi & 22  \iffalse 21.678088 +- 0.283 \fi & 0  \iffalse 0.18879568 +- 0.00637 \fi & 1400  \iffalse 1449.2984*0.9729 +- 1.97 \fi \\ 
\hline 
\multirow{3}{4em}{3000} 
& $q_{min}$=0.1 & 4200  \iffalse 4193.7804 +- 7.8 \fi & 1900  \iffalse 1860.148 +- 9.8 \fi & 260  \iffalse 260.93631 +- 16.9 \fi & 3700  \iffalse 6314.8647*0.59 +- 21.1 \fi \\ 
& $q_{min}$=0.5 & 4200  \iffalse 4211.7267 +- 7.8 \fi & 1500  \iffalse 1462.4432 +- 8.37 \fi & 220  \iffalse 218.24024 +- 18.6 \fi & 3900  \iffalse 5892.4101*0.66425 +- 21.8 \fi \\ 
& $q_{min}$=1 & 4200  \iffalse 4191.9015 +- 7.48 \fi & 1300  \iffalse 1258.9714 +- 7.08 \fi & 140  \iffalse 142.35464 +- 6.27 \fi & 3900  \iffalse 5593.2276*0.6928 +- 12.1 \fi \\ 
& $q_{min}$=5 & 4200  \iffalse 4191.6303 +- 7.41 \fi & 850  \iffalse 847.72088 +- 5.03 \fi & 66  \iffalse 65.634169 +- 3.19 \fi & 3900  \iffalse 5104.9854*0.76445 +- 9.5 \fi \\ 
& $q_{min}$=10 & 4200  \iffalse 4189.2645 +- 7.63 \fi & 680  \iffalse 678.69216 +- 3.81 \fi & 31  \iffalse 31.193611 +- 2.79 \fi & 3900  \iffalse 4899.1502*0.80355 +- 8.97 \fi \\ 
& $q_{min}$=50 & 4000  \iffalse 3958.6044 +- 7.65 \fi & 240  \iffalse 243.85393 +- 1.47 \fi & 2  \iffalse 2.4633981 +- 0.274 \fi & 3700  \iffalse 4204.9218*0.8904 +- 7.8 \fi \\ 
\hline 
\end{tabular} 

}
\caption{Cross section values for different merging parameter
  $q_{min}$ and different multiplicities of photons
  included in the matrix element calculations in (radiative)
  $\PGn\PAGn$ events with the 'hard photon'
  requirement.
  Number of significant digits indicates the statistical
    precision resulting from \whizard integration.
}
\label{tab:vv_tab_qscan_hard}
\end{table}

\begin{figure}[tbp]
  \includegraphics[width=0.49\textwidth]{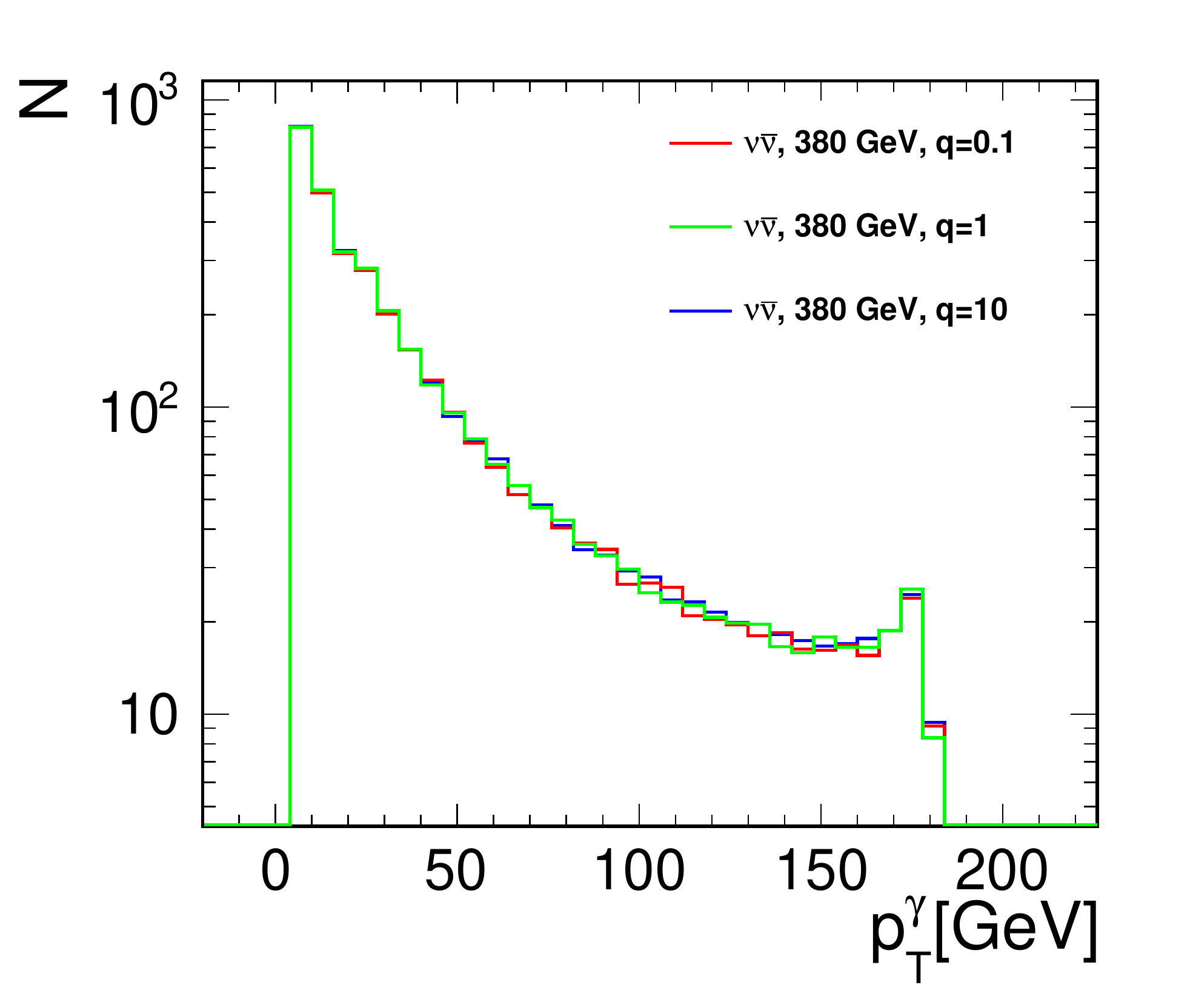}
  \includegraphics[width=0.49\textwidth]{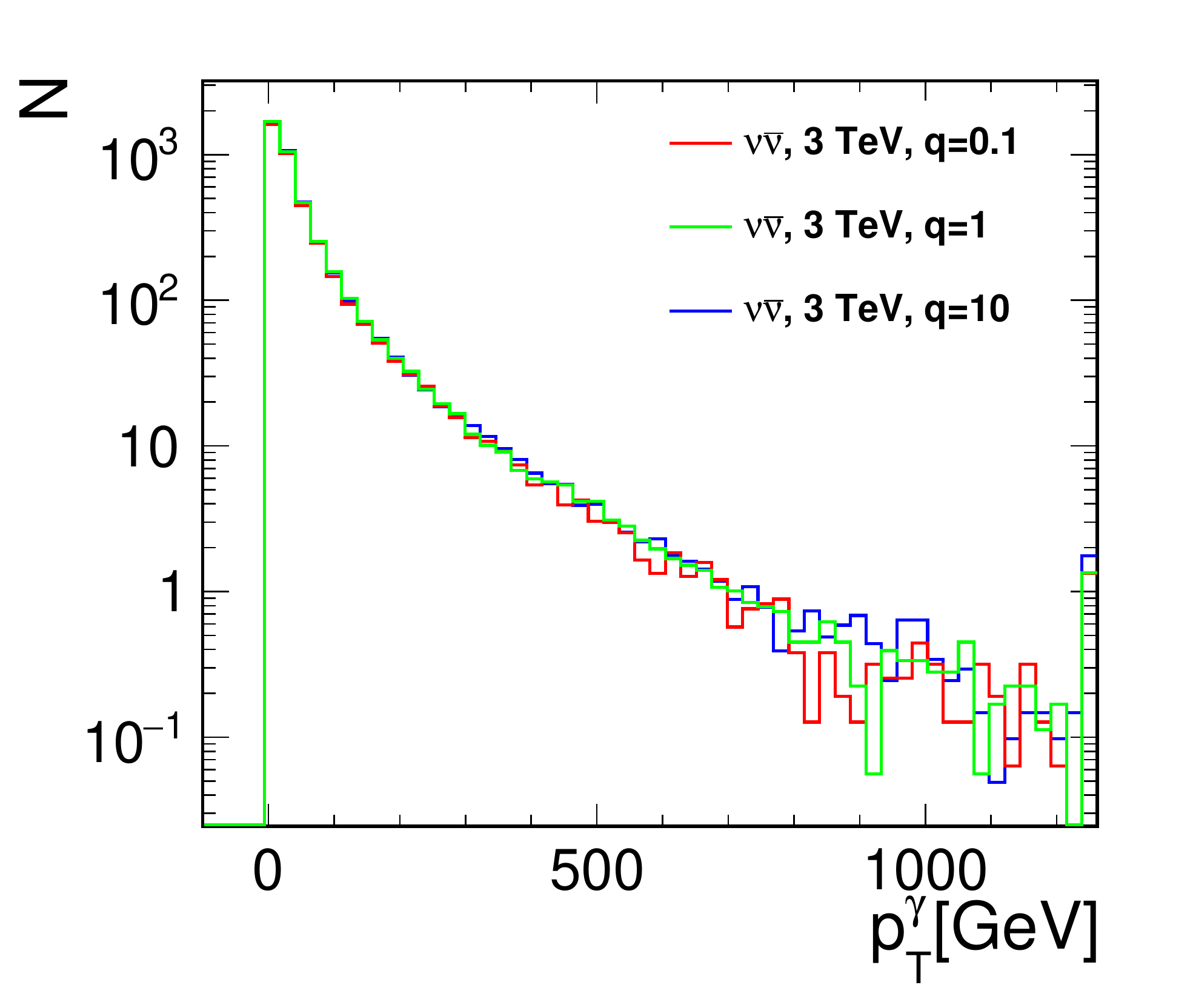}
  \caption{Distribution of the hard photon transverse momenta for
    radiative neutrino pair production at 380\,GeV (left) and 3\,TeV
    (right) in \whizard, for different values of the merging parameter $q_{min}$.
    Distributions are normalised to the number of events expected for
    integrated luminosity of 1\,fb$^{-1}$. }
  \label{fig:vv_pt_qscan}
\end{figure}

Figure \ref{fig:vv_pt_qscan} shows the distribution of the hard photon
transverse momenta for radiative neutrino pair production at 380\,GeV
and 3\,TeV, for different values of the merging parameter $q_{min}$.
It demonstrates that the photon transverse momentum distribution,
after hard photon selection, is not sensitive to the choice of the
$q_{min}$ parameter.

%-----------------------------------------------------------------------

\subsection{Comparison of hard photon selection for Whizard and the \kkmc generators}

To verify the accuracy of the proposed ISR-ME merging procedure we
compare the results of \whizard 2.8.2 with those of the \kkmc 4.22 code
\cite{Jadach:1999vf,Jadach:2013aha}.
The \kkmc includes soft-photon resummation in Coherent Exclusive
Exponentiation (CEEX)\cite{Jadach:2000ir}, based on the Yennie-Frautschi-Suura scheme
\cite{Yennie:1961ad} and the exact $\mathcal{O}(\text{Born}+\alpha^{3/2})$
real and virtual matrix element corrections. 
Therefore single photon differential distributions  in the \kkmc have a
full NLO accuracy. 
\whizard on the other hand includes all order resummation of soft and
soft-collinear, and hard-collinear emissions up to 3rd order in the
ISR.
The proposed procedure supplements \whizard's default precision with
exact hard matrix elements up
to order $\mathcal{O}(\text{Born}+\alpha^{3/2})$.
Therefore we expect that observables with at least 2 hard
non-collinear photons for processes without FSR should have a similar
accuracy as in the \kkmc while single photon ones will differ by missing
genuine 1-loop QED corrections.
%
%Below we investigate the numerical impact of those differences.
%
To allow for direct comparison of the two approaches, electroweak
corrections not available in \whizard are also disabled in the \kkmc.
These corrections contribute to 2-3\% of the total cross section.

\begin{figure}[tbp]

  \includegraphics[width=0.49\textwidth]{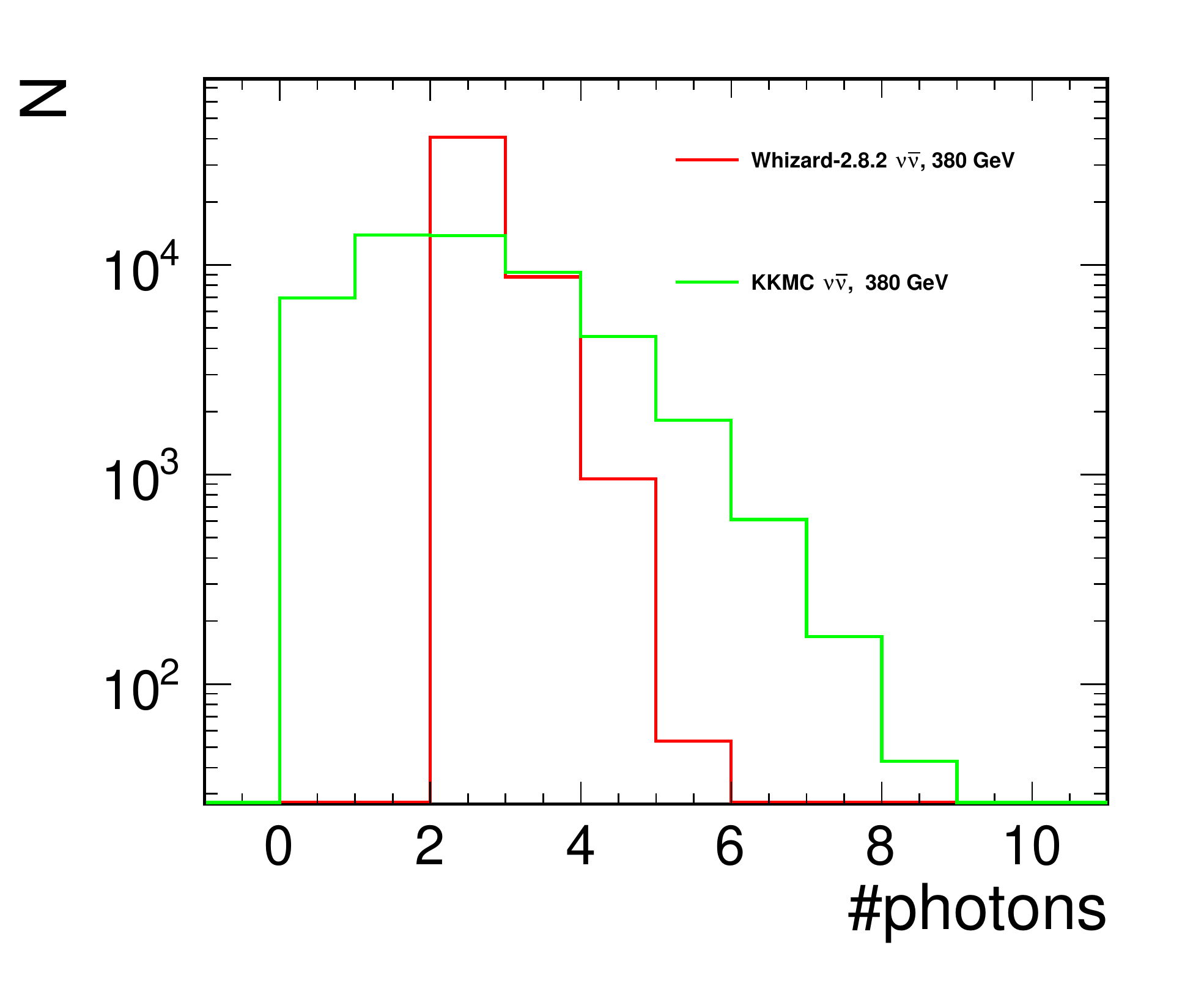}
  \includegraphics[width=0.49\textwidth]{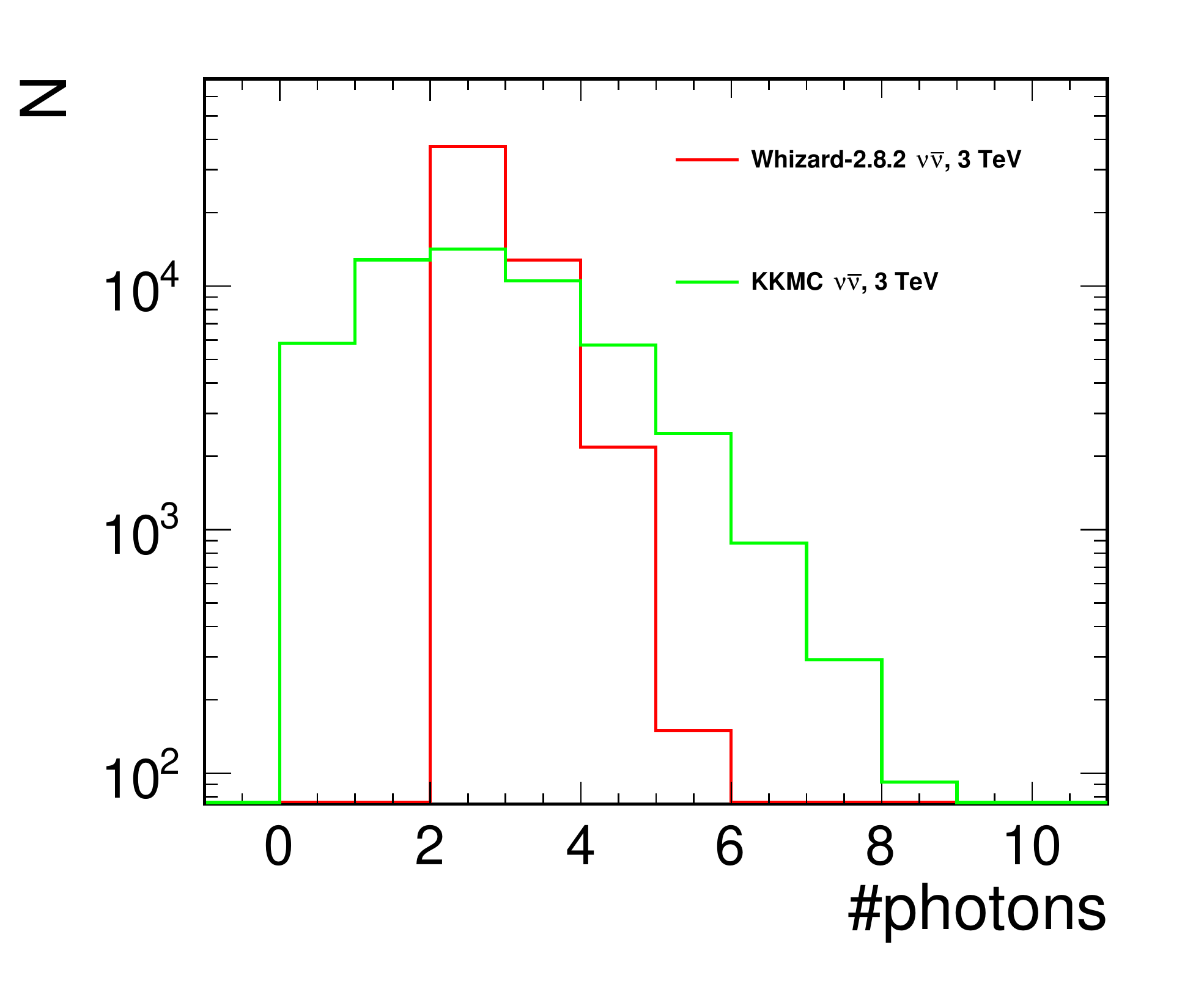}

  \includegraphics[width=0.49\textwidth]{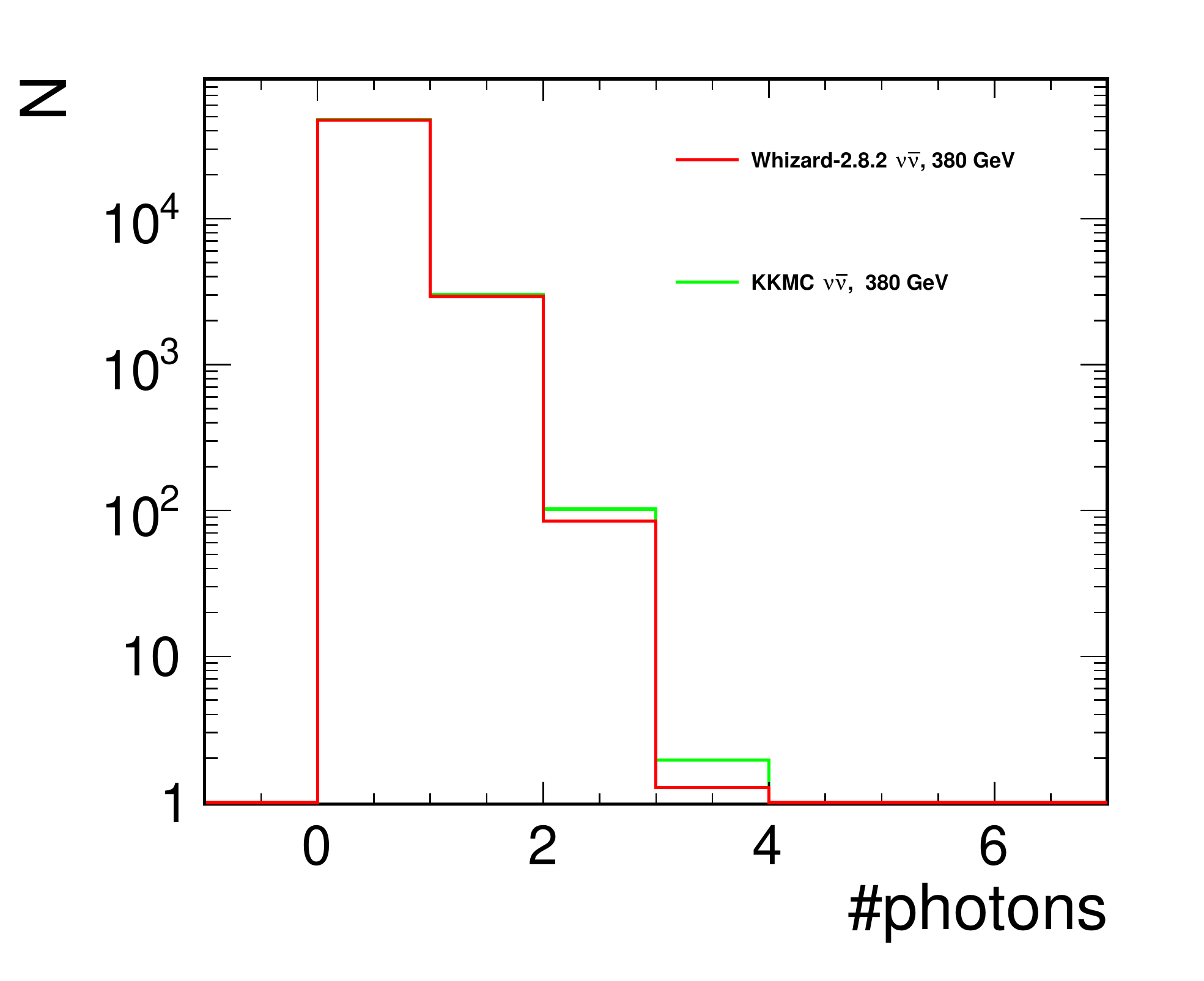}
  \includegraphics[width=0.49\textwidth]{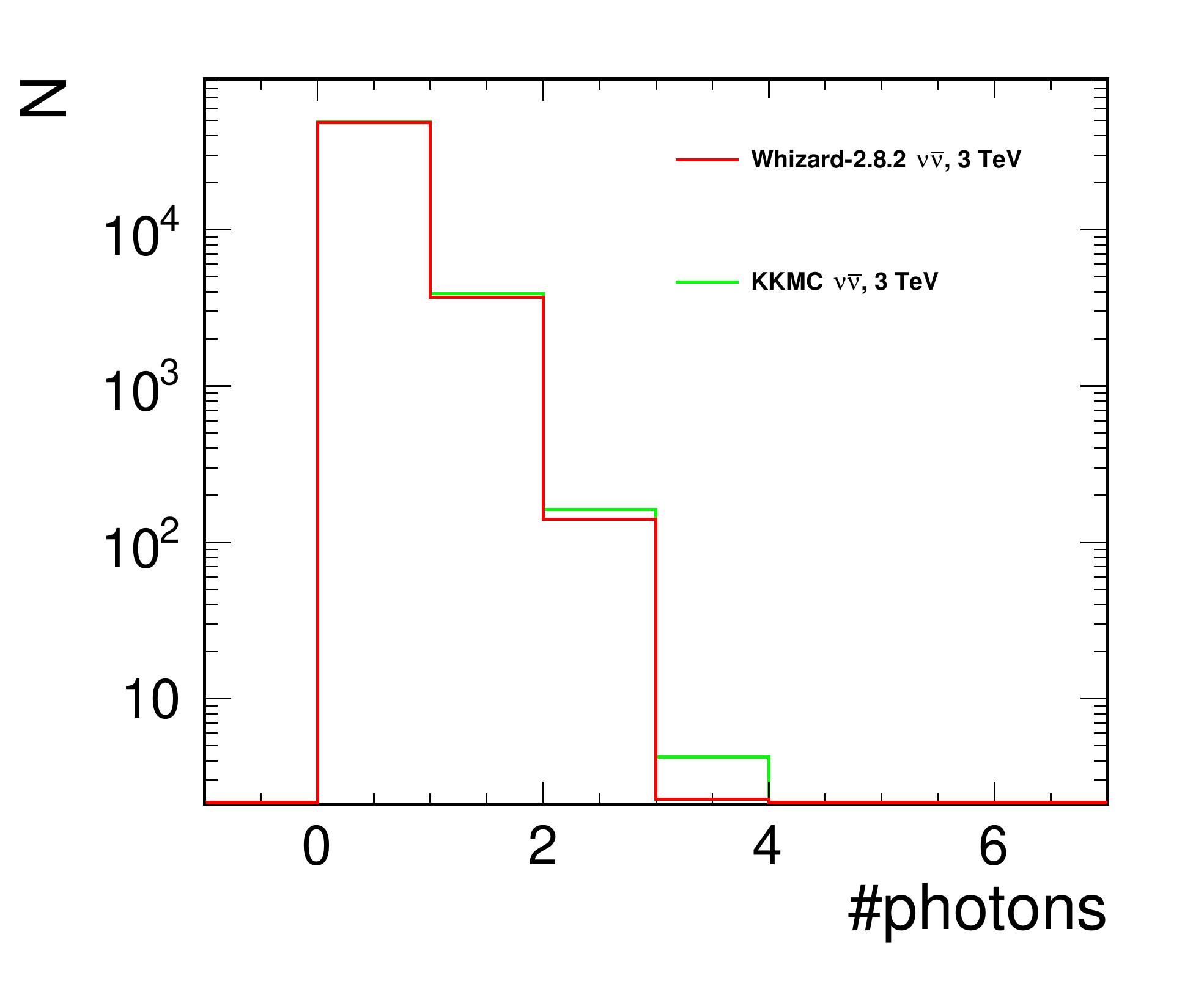}

  \caption{Distributions of the number of photons in the neutrino pair
    production events generated by \whizard and the \kkmc.
    Compared are multiplicities of all generated photons (upper row)
    and for those remaining after hard photon selection (lower row),
    for collision energy of 380\,GeV (left) and 3\,TeV (right).
    Distributions are normalised to the number of events expected for
    integrated luminosity of 1\,fb$^{-1}$.
  }
  \label{fig:vv_nPhot_compr}
\end{figure}

Shown in Fig.~\ref{fig:vv_nPhot_compr} are distributions of the
generated photon multiplicities, for \whizard and the \kkmc.
Respective total cross sections for those samples are 50.6\,pb and
51.1\,pb for the 380\,GeV case, 52.4\,pb and 52.8\,pb for 3\,TeV
scenario, with errors of order of 0.1\%.
If all generated photons are considered (upper row in
Fig.~\ref{fig:vv_nPhot_compr}), very different distributions are
obtained for the two Monte Carlo codes.
The \kkmc generates events with variable number of photons. Significant
fraction of neutrino pair-production events is generated without any
photon radiation but there are also events with up to 8 photons
in the final state.
For \whizard, there are always at least two photons in the final
state, corresponding to the photons generated from ISR parametrisation
for electron and positron beam.
Fractions of events with two to five photons correspond
to the relative contributions of processes with zero to three photons
generated on the ME level.
The ME photons generated by \whizard need to pass ISR-ME
merging criteria and that is why their contribution decreases faster with
photon multiplicity than in the case of the \kkmc.

\begin{figure}[tbp]

  \includegraphics[width=0.49\textwidth]{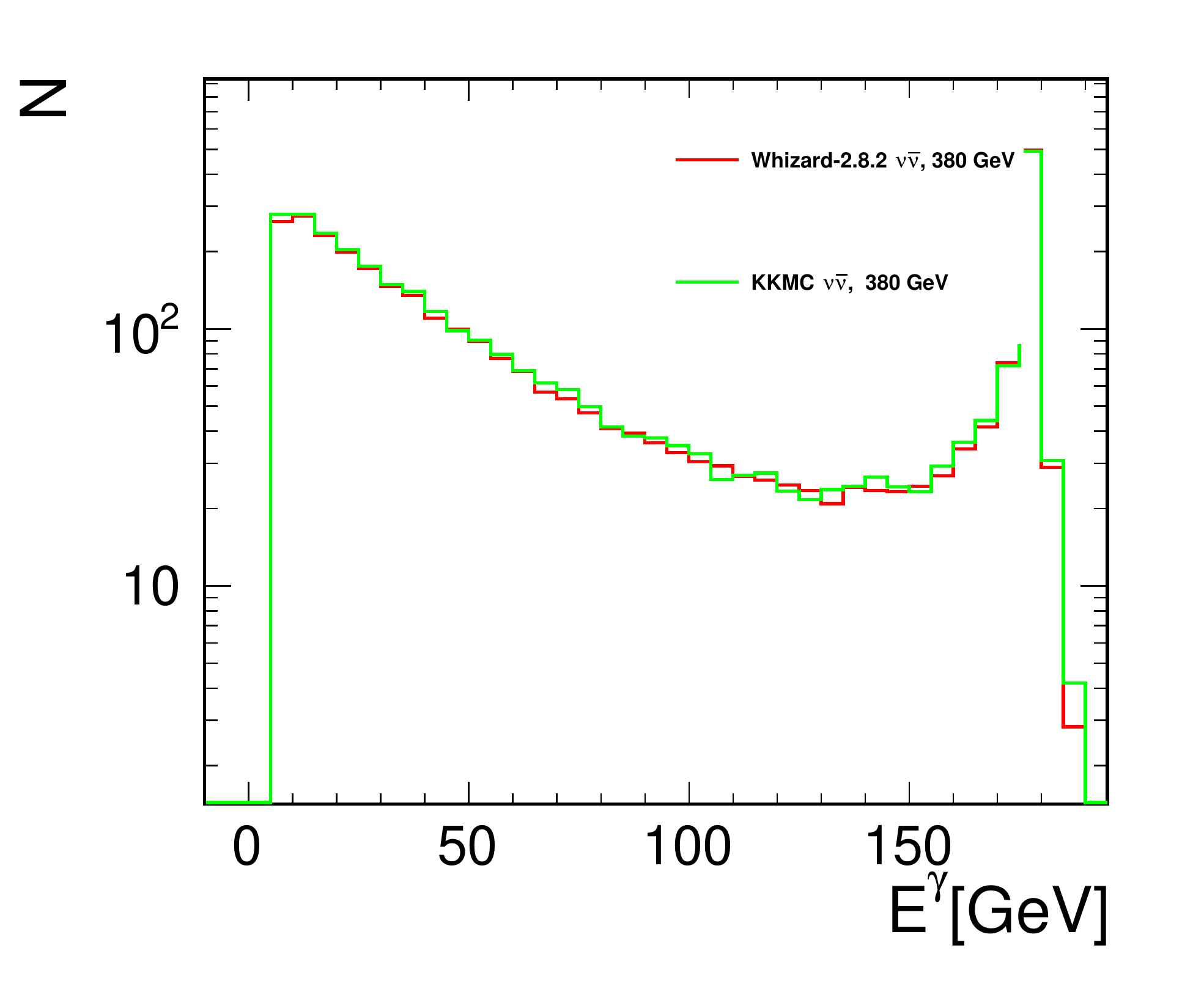}
  \includegraphics[width=0.49\textwidth]{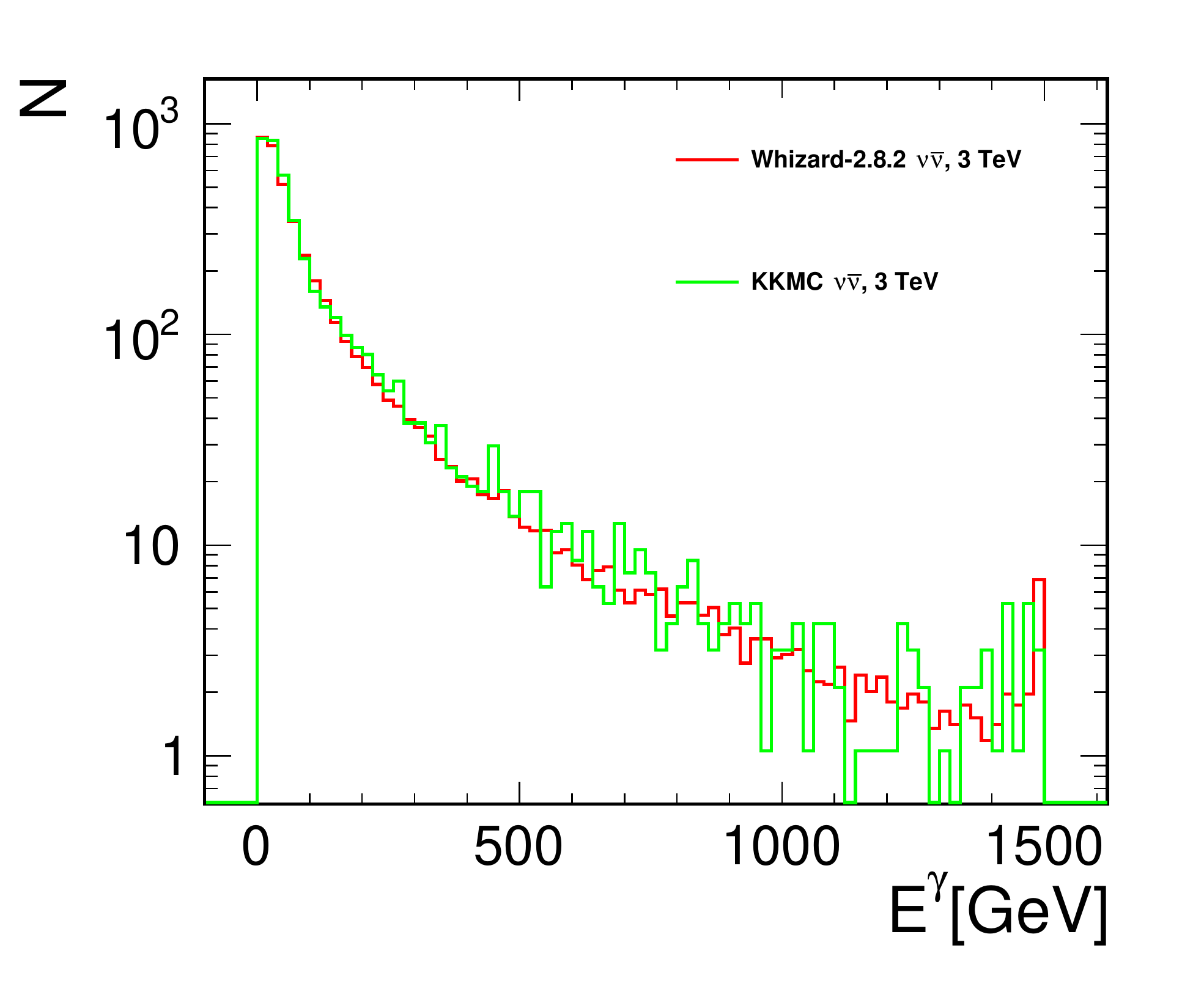}

  \includegraphics[width=0.49\textwidth]{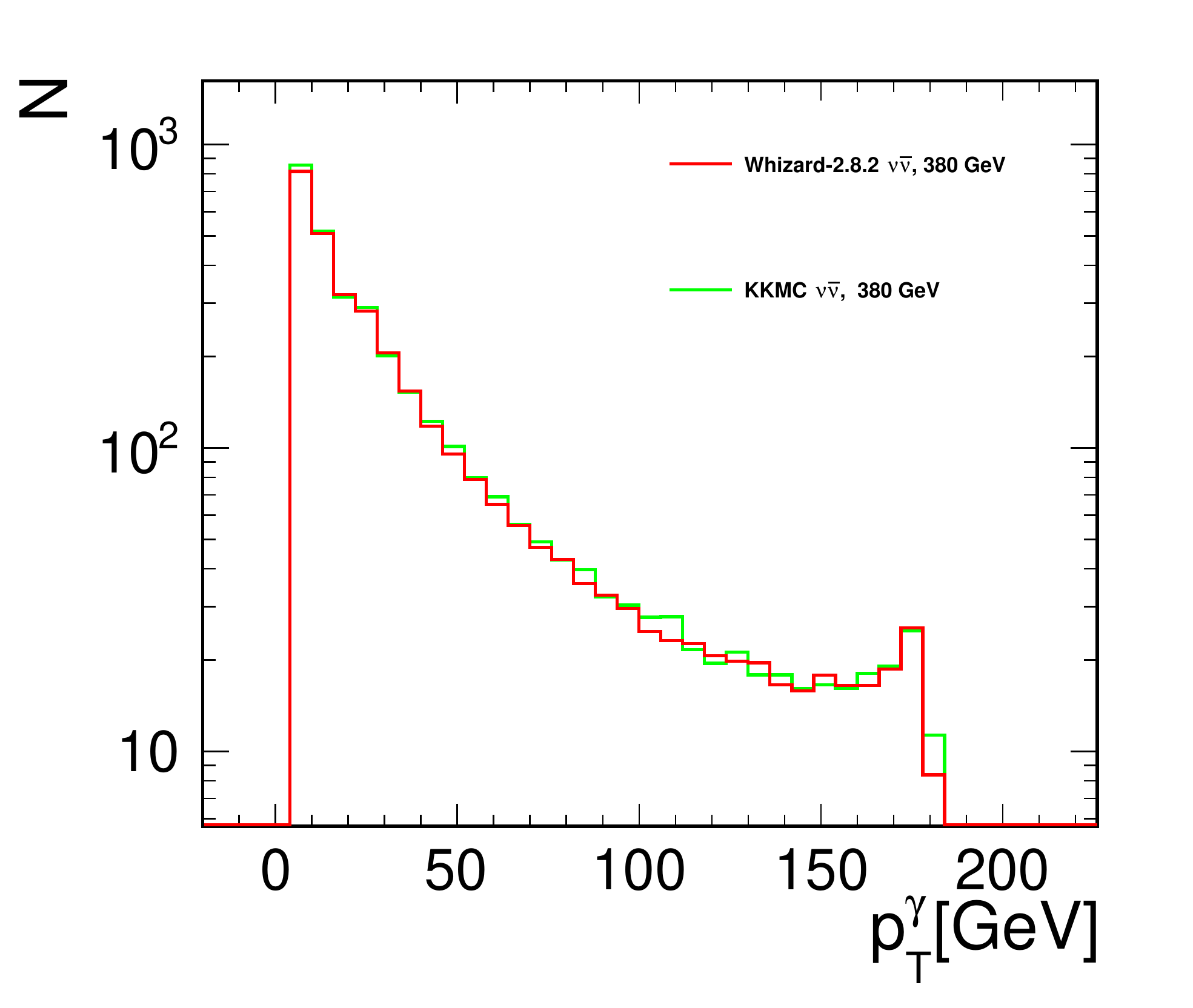}
  \includegraphics[width=0.49\textwidth]{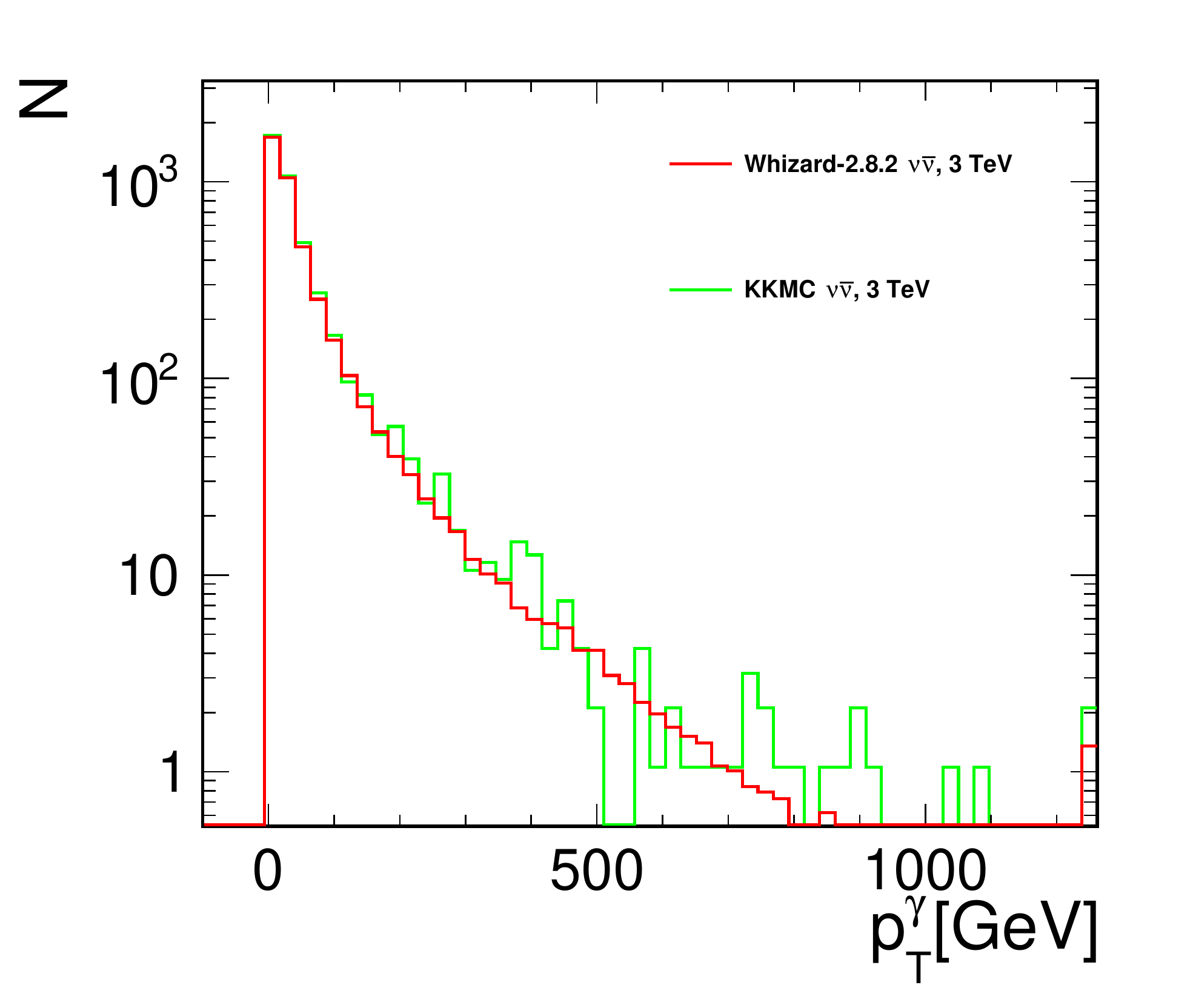}

  \caption{Distributions of the photon energy (upper row) and
    transverse momentum (lower row) in the neutrino pair
    production events generated by \whizard and the \kkmc,
    for collision energy of 380\,GeV (left) and 3\,TeV (right),
    after hard photon selection.
    Distributions are normalised to the number of events expected for
    integrated luminosity of 1\,fb$^{-1}$.
    }
  \label{fig:vv_kine_compr}
\end{figure}

However, after hard photon selection cuts are applied, corresponding
to the expected detector acceptance, the multiplicity distributions
obtained with \whizard and the \kkmc are in good agreement.
Very good agreement is also observed for photon energy and transverse
momentum distributions, which are compared in
Fig.~\ref{fig:vv_kine_compr}, for neutrino pair-production at 380\,GeV
and 3\,TeV.
Total cross sections for 380\,GeV samples are
3.1\,pb for both generators while for 3\,TeV scenario \whizard sample has
3.9\,pb and \kkmc 4.1\,pb.
This confirms that the proposed simulation procedure, including ISR-ME
merging gives a proper description of the hard photon production  in
neutrino pair-production events.

%
% We have also verified that the cross section for radiative neutrino
% pair-production at LEP2 obtained with the proposed procedure is in
% agreement with experimental results \cite{Achard:2003tx} when the same
% hard photon selection criteria are used.
%

% -*- mode: LaTeX; mode: auto-fill; mode: flyspell-prog; -*-

\section{Hard photons from Bhabha events}

When the Bhabha process is to be considered as the background source in the
mono-photon analysis, we should not set any constraints on the
final state leptons (electron or positron) on the generator level.
As they do not need to be observed in the detector, no requirement can
be imposed on the minimum momentum transfer or minimum lepton
scattering angle.
This is the main problem in generation of Bhabha events, as the cross
section for this process diverges for low scattering angles due to
the Coulomb singularity. 
However, most of the divergencies are removed when we require a hard
photon in the final state, as described in the previous section.
The only requirement which needs to be added, to remove collinear
divergencies, is the minimum angular separation between final state leptons
and photons.
Events with hard photon reconstructed in the detector, accompanied by
the charged lepton track can be easily rejected in the analysis, so
this cut should not result in any significant bias in the final
analysis results. 
Angular separation between photons and final state leptons greater than
$\theta^{e\gamma}_{min}=1^\circ$ was required for the results
presented in this section, see Appendix~\ref{sec:a2} for
an example of Sindarin steering file for \whizard.%
\footnote{
  While the choice of the $\theta^{e\gamma}_{min}$ value does
affect the numerical results presented in
Tabs.~\ref{tab:bha_tab_Escan} and \ref{tab:bha_tab_qscan} (cross
section after ISR rejection changes by about 20--30\% when changing the
separation cut by an order of magnitude), variations in results
obtained after electron rejection cuts are marginal.
% within the statistical uncertainties of the results. 
}

Shown in Tab.~\ref{tab:bha_tab_Escan} are the cross section values for
Bhabha scattering with emission of hard photons, for different
collision energies.
Compared are cross sections with up to three photons generated on the
ME level (at least one of them is required to pass the hard photon
selection) and the total cross section, after the ME-ISR merging
procedure (ISR rejection cuts), as described in the previous section.
Contrary to the neutrino pair-production, Bhabha scattering cross
section decreases fast with the increasing collision energy.

\begin{table}[tbp]
\centering
{\scriptsize
    \begin{tabular}{||c|c|c|c|c|c||}
\hline
\multirow{2}{4em}{$\sqrt{s}[GeV]$} & 
\multicolumn{3}{c|}{ Whizard-2.8 $\sigma(\Pep\Pem\rightarrow\Pep\Pem+N\gamma)$ [fb] } &
\multirow{2}{12em}{$\sigma(\Pep\Pem\rightarrow\Pep\Pem+N\gamma)$ [fb] after ISR rejection} \\ 
& $\Pep\Pem+\gamma_{ME}$ &  $\Pep\Pem+2\gamma_{ME}$ & $\Pep\Pem+3\gamma_{ME}$ & \\ 
\hline
%
%  240   235929     26059     1481    220416 
% +/-       672       150       28       896 
 240   & 236000  &  26100 &   1500  & 220000 \\
%
%  250   224189     24908     1447    209190 
% +/-       627       155       33       845 
 250   & 224000  &  24900 &   1400  & 209000 \\
%
%  380   140468     18422     1200    128213 
% +/-       437        83       30       539 
 380   & 140000  &  18400 &   1200  & 128000 \\
%
%  500   100213     14581     1065     89833 
% +/-       320        95       62       390 
 500   & 100000  &  14600 &   1100  &  89800 \\
%
% 1000    39789      7714      723     34446 
% +/-       146        41       27       159 
1000   &  39800  &   7700 &    700  &  34400 \\
%
% 1500    23029      5272      512     19419 
% +/-        93        36       26        98 
1500   &  23000  &   5300 &    500  &  19400 \\
%
% 3000     8838      2697      329      7194 
% +/-        43        32       17        47 
3000   &   8800  &   2700 &    300  &   7200 \\
\hline
\end{tabular} 

}

\caption{Cross section values for
  different collision energies and different multiplicities of photons
  included in matrix element calculations in Bhabha events with a requirement of at
  least one of the matrix elements photons to be 'hard'.
  Number of significant digits indicates the statistical
    precision resulting from \whizard integration.
}
\label{tab:bha_tab_Escan}
\end{table}

Compared in Tab.~\ref{tab:bha_tab_qscan} are the radiative Bhabha scattering
cross sections obtained for different merging scales $q_{min}$, for
two selected collision energies, 380\,GeV and 3\,TeV.
As before, the contributions of multiphoton events strongly depend on the
merging scale, while the total cross section after ISR rejection is
weekly dependent on $q_{min}$.
Only for $q_{min}=50$\,GeV (in particular at 380\,GeV) a significant
drop of the cross section is observed, as the merging requirement becomes
more restrictive than the hard photon selection (see Fig.~\ref{fig:q_plot}).
Otherwise, cross section values vary by up to 5\% for collision energy
of 380\,GeV and by up to 11\% for 3\,TeV, from the values obtained for
$q_{min}=1$\,GeV, when  $q_{min}$ is varied by an order of magnitude.
One has to note that the ISR approximation implemented in \whizard is
not stricly valid for the Bhabha scattering because of the Coulomb
singularity.
Cross section variation with  $q_{min}$ can also be attributed to the
contribution from the final 
state radiation (FSR) and the ISR-FSR interference, which are taken into
account in ME approach, but not in the ISR parametrisation.
When increasing the $q_{min}$ cut, the phase space
for photons generated with ISR parametrisation is increased while it
is reduced for ME generation.
This effect should be further reduced when additional selection cuts
(in particular electron veto) are applied in the analysis.

The influence of the beamstrahlung on the radiative Bhabha cross
section is significant, much larger than for the neutrino pair
production.
When CLIC luminosity spectra are taken into account, the radiative 
Bhabha cross section increases by 5\% at 380\,GeV and by over a factor
of 2  at 3\,TeV.   
This is due to the large beamstrahlung effects expected at 3\,TeV CLIC
and to the Bhabha scattering cross section decreasing rapidly with
energy.
The cross section contribution from the process with four ME photons
in the final state is found to be negligible.

\begin{table}[tbp]
\centering
{\scriptsize
    \begin{tabular}{||c|c|c|c|c|c||}
\hline
\multirow{2}{4em}{$\sqrt{s} [GeV]$} & \multirow{2}{4em}{$q_{min}[GeV]$} & 
	\multicolumn{3}{c|}{ Whizard-2.8 $\sigma(\Pep\Pem\rightarrow\Pep\Pem+N\gamma)$ [fb]} & 
	\multirow{2}{12em}{$\sigma(\Pep\Pem\rightarrow\Pep\Pem+N\gamma))$ [fb] after ISR rejection} \\ 
& & $\Pep\Pem+\gamma_{ME}$ &  $\Pep\Pem+2\gamma_{ME}$ & $\Pep\Pem+3\gamma_{ME}$ & \\ 
\hline
\multirow{3}{4em}{380}
% [140544.67, 29511.249, 3176.421, 173232.343637]
% [428.0, 138.0, 55.1]
% eff: 0.71 corr_xsec: 122956.9
& $q_{min}$=0.1 & 141000  & 29500 & 3200  & 123000 \\

% [140174.63, 21681.819, 1531.7769, 163388.221495]
% [412.0, 93.9, 36.5]
%  eff: 0.773 corr_xsec: 126294.2
& $q_{min}$=0.5 &  140000 & 21700 & 1500  &  126000 \\

%  [139973.16, 18423.43, 1246.9927, 159643.57931]
% [437.0, 83.4, 26.7]
% eff: 0.799 corr_xsec: 127561.6 
& $q_{min}$=1 &140000  &  18400  &  1200 & 128000  \\

% [141071.05, 9875.7128, 398.78347, 151345.549448]
% [475.0, 47.9, 4.73]
% eff: 0.867 corr_xsec: 131212.1 
& $q_{min}$=5 & 141000  & 10000 & 400 & 131000 \\

% [139841.92, 5089.2629, 165.39228, 145096.578955]
% [416.0, 41.8, 2.93]
%eff: 0.897 corr_xsec: 130192.3 
& $q_{min}$=10 & 140000 & 5100 & 170 & 145000 \\

% [11200.327, 161.99807, 2.0869674, 11364.412347]
% [38.4, 1.37, 0.0328]
% eff: 0.967 corr_xsec: 10989.2 
& $q_{min}$=50 & 11200 & 160 & 2 & 11000 \\

\hline 
\multirow{3}{4em}{3000}
% [8902.1444, 3669.1247, 417.1912, 12988.460217]
% [41.4, 22.7, 23.4
% eff: 0.494 corr_xsec: 6410.5 
& $q_{min}$=0.1 & 8900 & 3700 & 420 & 6400 \\

% [8958.0787, 3018.1153, 226.40942, 12202.603411]
% [51.5, 23.0, 12.0]
% eff: 0.571 corr_xsec: 6963.9 
& $q_{min}$=0.5 & 9000 & 3000 & 230 & 7000 \\

% [8860.8607, 2691.098, 276.93609, 11828.894754]
% [39.4, 16.2, 11.3]
% eff: 0.609 corr_xsec: 7199.5
& $q_{min}$=1 & 8800   & 2700 & 300 & 7200 \\ % ujednolicone do wynikow Filipa, z tab_Escan_bhabha.hard.tex, ma wieksze bledy

% [8871.5506, 1983.3129, 138.78733, 10993.650808]
% [47.0, 14.5, 4.03]
% eff: 0.692 corr_xsec: 7602.1 
& $q_{min}$=5 &  8900  & 2000  &  140 &  7600\\

% [8906.1758, 1337.4579, 94.475433, 10338.109091]
% [43.9, 5.93, 3.33]
% eff: 0.727 corr_xsec: 7513.8 
& $q_{min}$=10 & 8900  & 1300 & 90 & 7500 \\

% [7020.0952, 295.59411, 11.064267, 7326.753617]
% [29.5, 1.23, 0.149]
% eff: 0.829 corr_xsec: 6076.9 
& $q_{min}$=50 & 7000  & 300 & 10 &  6100\\

\hline 
\end{tabular} 

}

\caption{Cross section values for different merging parameter
  $q_{min}$ and different multiplicities of photons
  included in the matrix element calculations in Bhabha events with a requirement of at
  least one of the matrix elements photons to be 'hard'.
  Number of significant digits indicates the statistical
    precision resulting from \whizard integration.
}
\label{tab:bha_tab_qscan}

\end{table}

\begin{figure}[tbp]
  \includegraphics[width=0.49\textwidth]{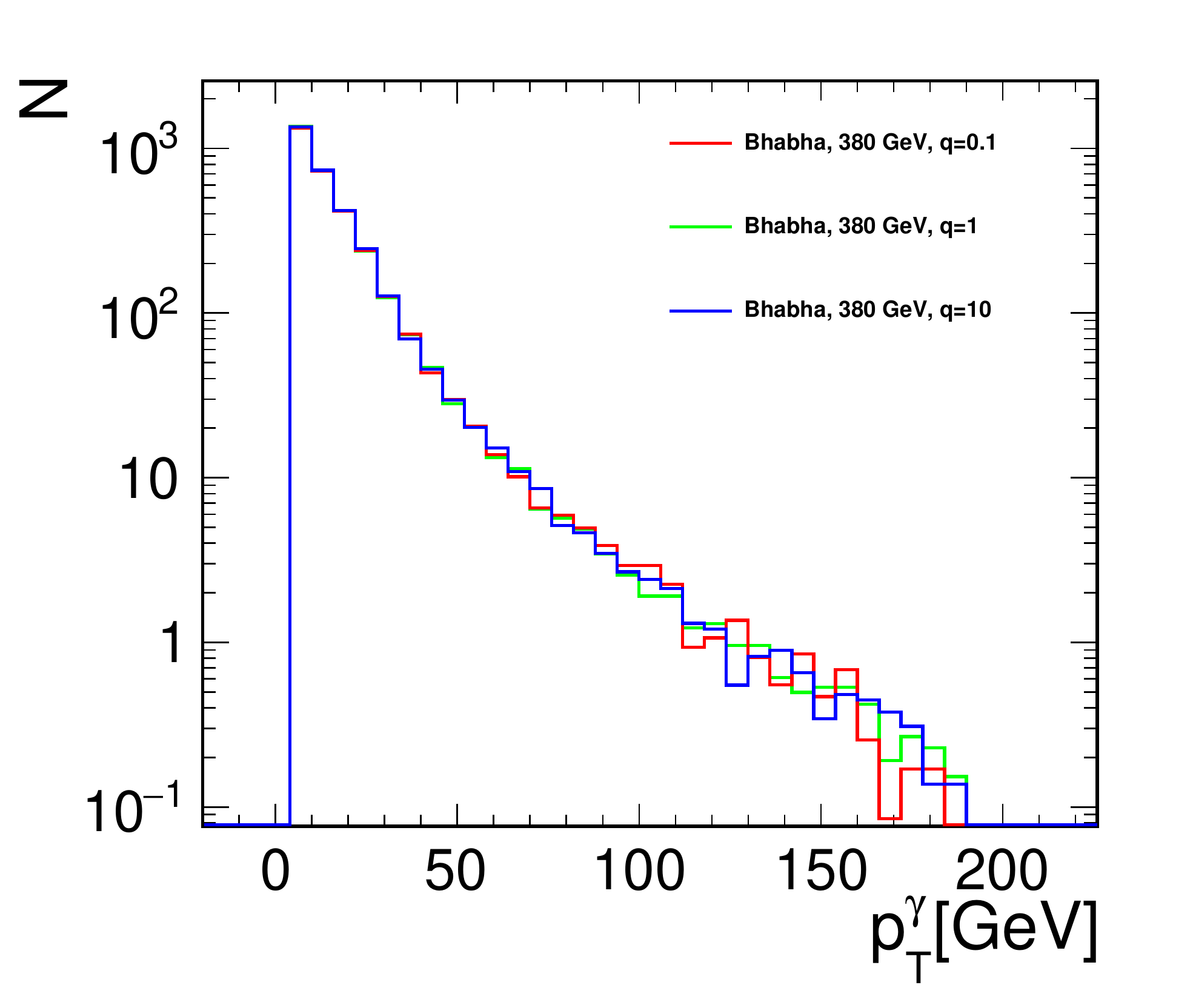}
  \includegraphics[width=0.49\textwidth]{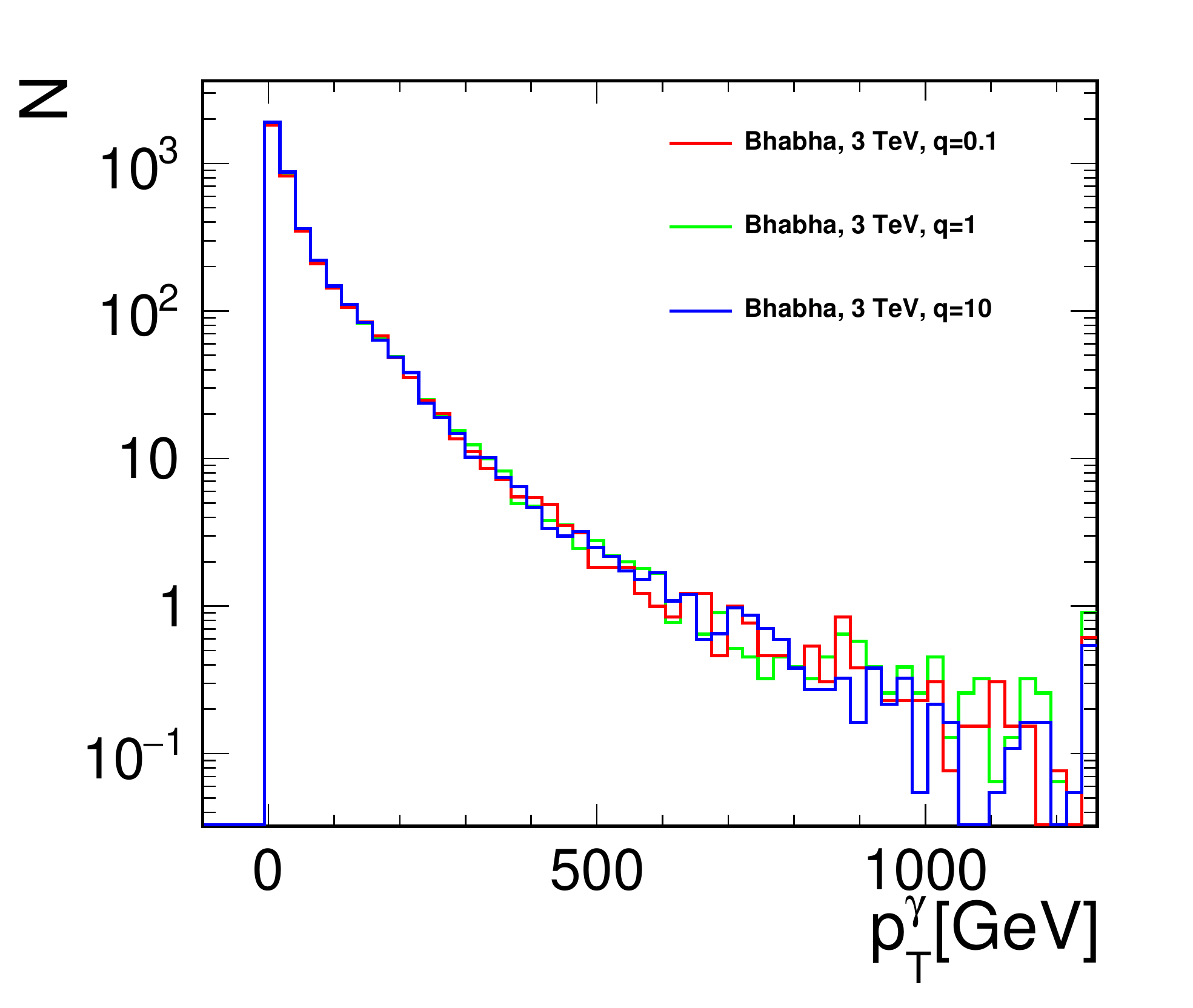}
  \caption{Distribution of the hard photon transverse momenta for
    radiative Bhabha events at 380\,GeV (left) and 3\,TeV
    (right) in \whizard, for different values of the merging parameter $q_{min}$.
    Distributions are normalised to the number of events expected for
    integrated luminosity of 1\,fb$^{-1}$.}
  \label{fig:bhabha_pt_qscan}
\end{figure}

Figure \ref{fig:bhabha_pt_qscan} shows the distribution of the hard photon
transverse momenta for radiative Bhabha events at 380\,GeV
and 3\,TeV, for different values of the merging parameter $q_{min}$.
Also in this case the photon transverse momentum distribution,
after hard photon selection, is not sensitive to the choice of the
$q_{min}$ parameter, confirming the validity of the proposed approach.

\begin{figure}[tbp]
  \centerline{\includegraphics[width=0.7\textwidth]{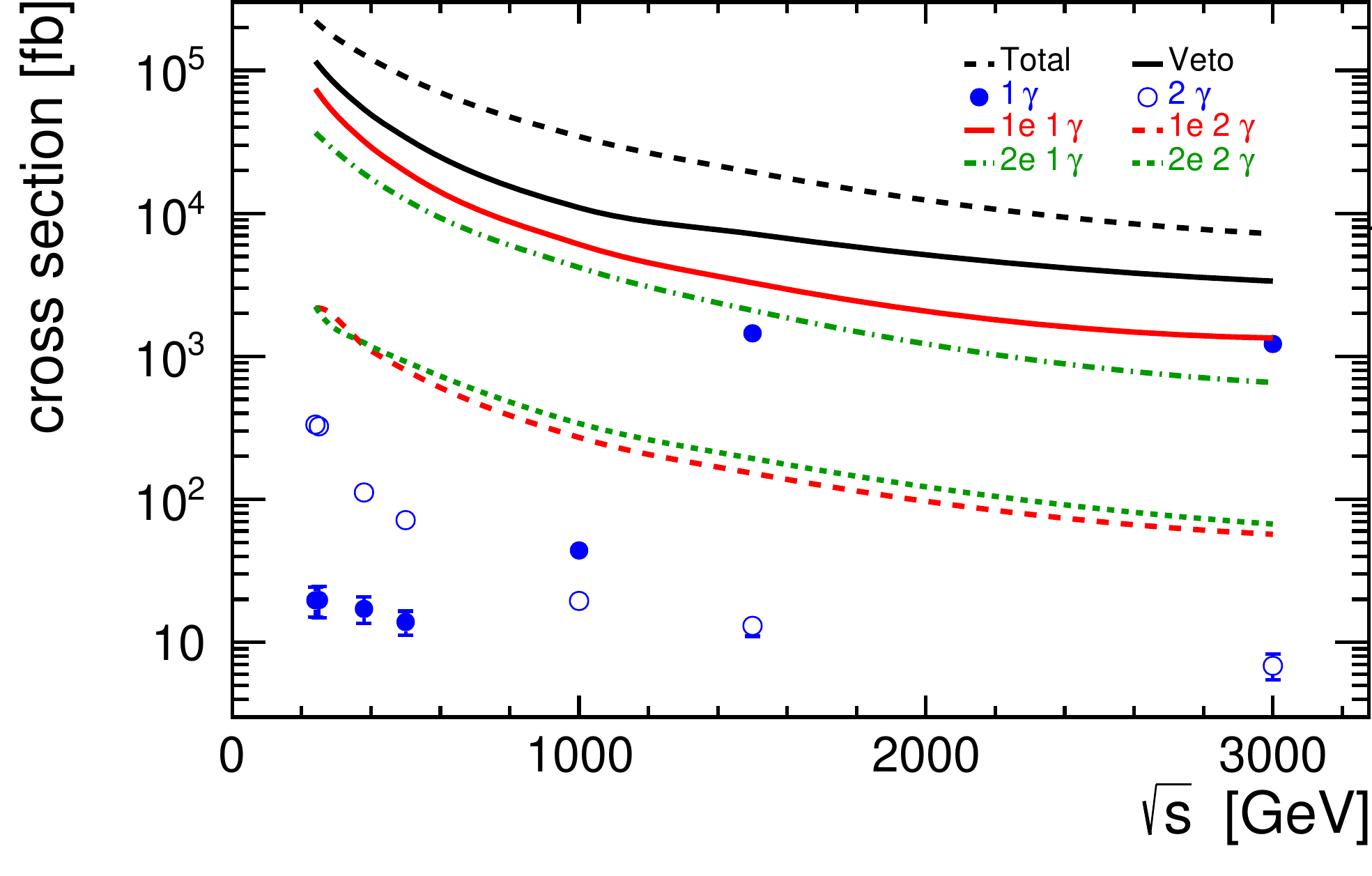}}
  \caption{Cross sections for radiative Bhabha scattering with at least one
    `hard' photon, after ISR rejection, as a function of the
    centre-of-mass energy.
    Cross sections for different final states observed in the central
    detector, calculated assuming no activity in the BeamCal or LumiCal detectors,
    are compared with the overall cross section obtained after
    applying BeamCal and LumiCal cuts (Veto)
    and the total cross section for the radiative Bhabha scattering
    (Total), as given in Tab.~\ref{tab:bha_tab_Escan}. 
  }
  \label{fig:bhabha_plot}
\end{figure}

While the cross section for neutrino pair production with hard photon
emission is dominated by events with only a single photon
visible in the detector, the situation is very different for the
radiative Bhabha scattering.
The transverse momentum of the hard photon has to be balanced either
by the transverse momentum of the scattered electron or positron, or
by emission of additional photons.
By applying a veto cut on the energy deposits in the BeamCal and
LumiCal detectors, below the acceptance region of the tracking
detectors (see Sec.~\ref{sec:detector}), about half of the radiative
Bhabha scattering events can be identified and rejected.\footnote{As
  the tracking information can not be used to distinguish between
  electrons and photons, the veto cut is also applied to additional ME
  photons produced in the BeamCal and LumiCal acceptance.}
For events passing such a veto cut, information from the tracking
detectors can be used to identify electrons and photons in the central
detector region.
Expected contributions from different final state topologies are presented in
Fig.~\ref{fig:bhabha_plot}. 
Samples are clearly dominated by events with scattered electron(s) in the
central region.
For energies up to 500\,GeV, Bhabha events with only a single photon
visible in the detector ($1\gamma$) contribute only to a tiny fraction
of the cross section, about 10--20\,fb.
Contribution of events with two photons in the final state is about an
order of magnitude higher.
Only for the centre-of-mass energy of 1\,TeV and above, mono-photon
event contribution from Bhabha scattering increases significantly.
This is due to the fact, that for the higher beam energy, electrons
can balance transverse momentum of the hard photon even when scattered
below the BeamCal acceptance.
This indicates that the minimum photon transverse momentum required in
the event selection should probably be increased at the high energy
running. 
Results presented in Fig.~\ref{fig:bhabha_plot} indicate also that the
detector performance, in particular performance of BeamCal and
LumiCal, as well as its modelling in the event simulation, is crucial
for the understanding and suppression of the Bhabha background.
Also, measurement of events with two reconstructed photons ($2\gamma$)
can give an important handle for verifying the detector performance
and Monte Carlo predictions. 

% -*- mode: LaTeX; mode: auto-fill; mode: flyspell-prog; -*-

\section{Conclusions}

Production of mono-photon events in $\Pep\Pem$ collisions is
considered as a possible signature for many BSM scenarios.
Analysis of the energy
spectrum and angular distributions of those photons can shed light
on the nature of new physics phenomena.
Therefore, the precise modelling of the mono-photon processes is
crucial.
In this paper the procedure has been presented which allows for a
reliable simulation of the mono-photon events in \whizard.
We focus on the two Standard Model background processes with the same
final state: radiative neutrino pair production and radiative Bhabha
scattering, and on the energy range considered for the future
$\Pep\Pem$ collider projects: from 240\,GeV for CEPC up to 3\,TeV for CLIC.
Presented are cross sections and kinematic
distributions of the mono-photon events, based on the proposed
merging of the matrix element calculations with the lepton ISR
structure function implemented in \whizard.
Results of such a procedure have been cross-checked with predictions
of the \kkmc generator for neutrino pair production processes.
The proposed procedure should be particularly useful for simulation of BSM
mono-photon events in \whizard.
The dominant background contribution is expected to result from the
radiative neutrino pair production events, while the background from
the Bhabha scattering will very strongly depend on the detector
performance, electron veto efficiency in particular.

\section*{Acknowledgements}

We thank
J.-J.\,Blaising,
S.\,Jadach,
W.\,Kilian,
J.\,Reuter
and
U.\,Schnoor
for valuable comments and suggestions.
This contribution was supported by the National Science Centre, Poland,
the OPUS project under contract UMO-2017/25/B/ST2/00496 (2018-2021) and
the HARMONIA project under contract UMO-2015/\-18/M/ST2/00518 (2016-2020).

\clearpage

\section*{Appendix}

\renewcommand{\thesubsection}{A.\arabic{subsection}}

\xspace

\subsection{Sindarin file for generation of radiative $\Pep\Pem\!\!\to\PGn\PAGn$ events}

\label{sec:a1}

\lstinputlisting[firstline=1, lastline=1000,%
  caption=Sindarin file for generation of $\PGn\PAGn(\PGg)$ background]%
                {tex/vv_append.sin}

\subsection{Sindarin file for generation of radiative Bhabha events}

\label{sec:a2}

\lstinputlisting[firstline=1, lastline=1000,%
  caption=Sindarin file for generation of $\Pep\Pem(\PGg)$ background]%
                {tex/bhabha_append.sin}

% add references
\printbibliography[title=References]

\end{document}